\documentclass{aa}  

\usepackage{float}
\usepackage{graphicx}
\usepackage{xcolor}
\usepackage{amsmath}	
\usepackage{amssymb}	
\usepackage{txfonts}
\usepackage[pdfauthor={Kristina Monsch, Giovanni Picogna, Barbara Ercolano, Thomas Preibisch},
pdftitle={The imprint of X-ray photoevaporation of planet-forming discs on the orbital distribution of giant planets -- II. Theoretical predictions},
pdfsubject={},
pdfkeywords={planet--disk interaction, planet formation, accretion disks, numerical hydrodynamics}]{hyperref} 
\usepackage{graphicx}  
\graphicspath{{plots/}} 

\newcommand\Msun{M_\odot}
\newcommand\Mjup{M_\mathrm{J}}
\newcommand\Mpl{M_{\rm p}}
\newcommand\Mdisc{M_\mathrm{d}}
\newcommand\Lx{L_\mathrm{X}}

\newcommand\yr{\mathrm{yr}}
\newcommand\Myr{\mathrm{Myr}}
\newcommand\Gyr{\mathrm{Gyr}}

\newcommand\ergs{\mathrm{erg\,s^{-1}}}
\newcommand\au{\mathrm{au}}

\newcommand\gcm{\mathrm{g\,cm^{-2}}}
\newcommand\Msunyr{M_\odot\,\mathrm{yr^{-1}}}
\newcommand\tclear{t_\mathrm{c}}
\newcommand\tform{t_\mathrm{form}}
\newcommand\Mdotwind{\dot{M}_\mathrm{w}}
\newcommand\Sigmadotwind{\dot{\Sigma}_\mathrm{w}}
\newcommand\Rgap{R_{\mathrm{gap}}}
\newcommand\tmigr{t_{\mathrm{m}}}
\newcommand\tnu{t_{\mathrm{\nu}}}

\newcommand\spock{\texttt{SPOCK}}

\newcommand\kepler{\textit{Kepler}}

\defcitealias{Picogna+2019}{P19}
\defcitealias{Owen+2012}{O12}

\begin{document} 

   \titlerunning{Disc dispersal \& the orbital distribution of giant planets}
   \authorrunning{Monsch et al.}
   
   \title{The imprint of X-ray photoevaporation of planet-forming discs on the orbital distribution of giant planets -- II. Theoretical predictions}

   \author{Kristina Monsch \inst{1,3}
          \and
          Giovanni Picogna \inst{1}
          \and
          Barbara Ercolano \inst{1,2}
          \and
          Thomas Preibisch \inst{1}
          }

   \institute{Universit\"ats-Sternwarte, Ludwig-Maximilians-Universit\"at M\"unchen, Scheinerstr.~1, 81679 M\"unchen, Germany\\
              \email{[monsch, picogna, ercolano, preibisch]@usm.lmu.de}
        \and
            Exzellenzcluster 'Origins', Boltzmannstr.~2, 85748 Garching, Germany 
        \and 
            Smithsonian Astrophysical Observatory, 60 Garden Street, Cambridge MA02138, USA 
             }

   \date{Received; accepted }

 
  \abstract
   {Numerical models have shown that disc dispersal via internal photoevaporation driven by the host star can successfully reproduce the observed pile-up of warm Jupiters near 1--$2\,\au$. However, since a range of different mechanisms have been proposed to cause the same feature, clear observational diagnostics of disc dispersal leaving an imprint in the observed distribution of giant planets could help in constraining the dominant mechanisms.}
   {We aim to assess the impact of disc dispersal via X-ray-driven photoevaporation (XPE) on giant planet separations in order to provide theoretical constraints on the location and size of any possible features related to this process within the observed semi-major axis distribution of giant planets.}
   {For this purpose, we perform a set of 1D planet population syntheses with varying initial conditions and correlate the gas giants' final parking locations with the X-ray luminosities of their host stars in order to quantify observables of this process within the semi-major axis versus host star X-ray luminosity plane of these systems.}
   {We find that XPE does create an under-density of gas giants near the gravitational radius, with corresponding pile-ups inside and/or outside this location. However, the size and location of these features are strongly dependent on the choice of initial conditions in our model, such as the assumed formation location of the planets. }
   {XPE can strongly affect the migration process of giant planets and leave potentially observable signatures within the observed orbital separations of giant planets. However, due to the simplistic approach employed in our model, which lacks a self-consistent treatment of planet formation within an evolving disc, a quantitative analysis of the final planet population orbits is not possible. Our results, however, should strongly motivate future studies to include realistic disc dispersal mechanisms in global planet population synthesis models with self-consistent planet formation modules.}


   \maketitle

\section{Introduction}
\label{sec:introduction}

Planet population synthesis models have become a well-established tool for directly comparing observational data with theoretical models of planet formation and evolution. 
Such models are based on the simple assumption that the observed diversity of extrasolar planets stems from the diversity in initial conditions in the nurseries of planetary systems, the planet-forming discs. 
By stochastically varying system parameters, such as the disc mass, the dust-to-gas ratio, or the amount of planetary embryos within a system, one can directly infer and isolate the impact of specific physical processes on the overall planet formation efficiency and ultimately predict the properties of planetary systems using an ensemble of statistically independent models. 
By unifying as many physical processes as possible, such models therefore provide a direct link between the observed population of planetary systems and the predictions of theoretical models \citep[see e.g.][for detailed reviews]{Benz+2014, Mordasini2018} 

Due to the large number of unknowns during the formation process of planets, there exists a variety of planet population synthesis models, which are centred on different aspects of the exoplanet demographics and therefore have different degrees of complexity.
The most comprehensive ones connect the earliest stages of planet formation with the later dynamical evolution of the fully formed planets, long after the gas disc has been dissipated \citep[e.g.][]{IdaLin2004a, Thommes+2008, Mordasini+2009, HellaryNelson2012, ColemanNelson2014, Bitsch+2015b, Ronco+2017, Ida+2018, Forgan+2018, Emsenhuber+2020}.
Based on such global approaches, some studies focus specifically on individual processes, such as the impact of pebble accretion onto planet formation and planetary system architectures \citep[e.g.][]{Bitsch+2019, Ndugu+2018, Ndugu+2019, Izidoro+2019} or the importance of the early infall phase from the parent molecular cloud core in setting the initial properties of protoplanetary discs \citep[e.g.][]{Schib+2020}. 

However, due to the complexity of these calculations, planet population synthesis models include by necessity simplified treatments of more complicated physical processes.
They typically adopt 1D parameterisations for disc and planet evolution, which are usually derived from more complex, multi-dimensional hydrodynamical calculations. Often, however, these prescriptions neglect subtler effects, or they may be only applicable to a specific subset of the modelled parameter space. 
One example is the so-called impulse approximation \citep{LinPapaloizou1979, LinPapaloizou1986}, which is commonly used in 1D models to calculate the torques exerted onto planets embedded in a gaseous disc. 
While it yields reasonable results for most disc-planet systems, \citet{Monsch+2021} show that it fails to correctly describe the migration rates of gas giants in transition discs with evacuated inner cavities. 
While the impulse approximation would predict an accelerated inward migration of giant planets in such systems, these authors have shown using 2D \texttt{FARGO} simulations that planet migration should cease as soon as the disc inside the planetary orbit is depleted of gas.
The reason for this is the formation of strongly positive torques right at the gap edge, which act as a planet trap, preventing any further inward migration of the planet.
This has important consequences for the final orbital location of giant planets and should therefore be taken into account in 1D models when calculating the migration tracks of giant planets in evolving protoplanetary discs.

This illustrates the Achilles heel of planet population synthesis models as their final outcome relies heavily on the accuracy of the employed prescriptions. For example, many models lack a detailed treatment of disc dispersal via photoevaporative winds, which are, however, a crucial ingredient in theoretical models for reproducing the observed disc lifetimes and strongly affect the final accreted gas mass of giant planets in a simulation.
Models often include a combination of internal and external photoevaporation via extreme (EUV) and far-ultraviolet (FUV) photons.
However, it has been shown that stellar EUV photons impinging on the circumstellar disc are already readily absorbed by small columns of neutral hydrogen and should therefore barely penetrate into the disc \citep[e.g.][]{Alexander+2004} and thus barely contribute to its heating and ionisation at larger radii. 
Photons fluxes of $\Phi \gtrsim 10^{41}\,\mathrm{s^{-1}}$ would be required to produce mass loss rates of $10^{-10}\,\Msun\,\yr^{-1}$ \citep[e.g.][]{Alexander+2006b, AA09}. However, observations of photoevaporative winds traced by free-free emission hint towards EUV fluxes being too low to dominate the dispersal of the discs around young T~Tauri stars \citep{Pascucci+2014, Macias+2016}; consequently by assuming purely EUV-driven winds, the impact of photoevaporation on dispersing circumstellar discs is likely to be underestimated (see however \citet{WangGoodman2017} and \citet{Nakatani+2018}. who come to a different conclusion).
Further, the importance of internal FUV-dominated photoevaporation on driving disc evolution is still matter of debate. While thermochemical models suggest significantly increased mass loss rates compared to the pure-EUV models \citep{Gorti+2009, GortiHollenbach2009}, these results still need to be confirmed by future hydrodynamical calculations.
On the other hand, external photoevaporation models driven by EUV and FUV photons by nearby high-mass stars have grown in importance in recent years \citep{Matsuyama+2003a, Winter+2018, Winter+2020}. In contrast to the internal models, detailed radiation-hydrodynamical calculations for external photoevaporation do exist \citep{Facchini+2016, Haworth+2016, Haworth+2018} but are not employed in current planet population synthesis approaches.

\section{The impact of disc dispersal on planetary orbits}
\label{sec:intro_PE}

Photoevaporative disc clearing has been suggested to have a dramatic impact on the semi-major axis distribution of giant planets \citep[e.g.][]{Matsuyama+2003b, HasegawaPudritz2012}, and recent numerical efforts have shown that photoevaporation by EUV and/or X-ray photons can indeed reproduce the observed pile-up of Jupiter-mass planets close to 1--$2\,\au$ \citep[\citealt{AP12, ER15}; however, see also][]{WiseDodsonRobinson2018}.
By heating the gas in the surface layers of the disc, the gas becomes unbound beyond the so-called gravitational radius:

\begin{equation}
\label{eq:Rg}
    R_\mathrm{g}= \frac{GM_\star}{c_\mathrm{s}^2} \approx 8.9\,\au \left( \frac{T_\mathrm{gas}}{10^4\,\mathrm{K}} \right)^{-1} \left( \frac{M_\star}{\Msun} \right),
\end{equation}
where $G$ is the gravitational constant, $M_\star$ the stellar mass, $c_\mathrm{s}$ the sound speed, and $T_\mathrm{gas}$ is the temperature of the heated gas layer \citep{Owen+2012}.
For X-ray-driven photoevaporation (XPE) of a $0.7\,\Msun$ star, $T_\mathrm{gas} \approx 10^3$--$10^4\,\mathrm{K}$, so that $R_\mathrm{g} \approx6$--$60\,\au$. This produces centrifugally launched, pressure-driven disc winds which result in the opening of an annular, gas-free gap inside of $R_\mathrm{g}$, fully decoupling the inner from the outer disc \citep[see][for a review]{Alexander+2014}.
Photoevaporation can therefore naturally provide a parking radius for inward migrating planets as their migration (which is a result of the angular momentum exchange between the planet and the gas-parcels in the disc) is ultimately stopped, once they reach the gas-free cavity. 
Planets inwards of the gap continue migrating shortly, while the inner disc is viscously drained, leading to a pile-up of planets just inside the photoevaporative gap located at the gap-opening radius (or `critical radius'\footnote{Analytic models have shown that there can already be significant mass loss due to photoevaporation already starting at radii of $R_\mathrm{crit} \approx 0.1$--$0.2\,R_\mathrm{g}$ \citep{Liffman2003, Adams+2004, Font+2004, Dullemond+2007_PP5}, while earlier models predicted that all gas would be fully gravitationally bound within $R_\mathrm{g}$.}, henceforth  $R_\mathrm{gap}$).
Planets outside of $R_\mathrm{gap}$ also continue migrating inwards; however, they are at the latest stopped once they reach the expanding photoevaporative gap. 
Compared to pure EUV-models \citep{AP12}, XPE has been shown to be even more effective in reproducing the observed pile-up of giants close to 1--$2\,\au$ \citep{ER15}. As the mass loss is more extended, it can park a larger fraction of planets, especially the more massive ones, at larger radii. 

However, other models have also been proposed as possible cause for the pile-up of gas giants near $1\,\au$, such as a reduction of type~II migration rates  \citep[e.g.][]{Ida+2018} or magnetically driven disc winds that can generate pile-ups within the surface density \citep{Suzuki+2016, Chambers2019}. 
Consequently, clear diagnostics for XPE shaping giant planet architectures are needed for differentiating between the possible driving mechanisms. 
\citet{Monsch+2019} have investigated the possibility that disc dispersal via XPE may leave such an observable imprint. By self-consistently calculating the X-ray luminosities, $\Lx$, of giant planet-hosting stars and correlating them with the semi-major axes, $a$, of their planets, they found a suggestive void within the $\Lx$--$a$-plane, which may hint towards XPE parking the planets close to the photoevaporative gap. However, due to the limited amount of X-ray observations, they could not prove the statistical significance of this void without either increasing the sample size drastically or having an accurate theoretical model that would predict the exact location and size of this gap a priori. 

Motivated by the observational study presented in \citet{Monsch+2019}, we aim to use theoretical models to look for possible signatures of XPE in the observed semi-major axis distribution of giant planets. 
For this purpose, we performed detailed 1D planet population synthesis models including giant planet migration and disc dispersal via internal photoevaporation driven by the host star. By varying key system parameters, such as the stellar X-ray luminosity, the planet mass and the planetary formation time, we predict what kind of features photoevaporation is expected to leave within the orbital distribution of gas giants. 
Our study is conceptually similar to previous work by \citet{Jennings+2018}, who have compared the impact of EUV, X-ray and FUV-photoevaporation on the orbital distribution of a given set of giant planets. 
Our study, in turn, solely focuses on XPE and intends to explore its impact on the migration process of giant planets as a possible origin of the over- and under-densities observed in the demographics of giant planets. 
We aim to provide a comprehensive model that can aid the interpretation of the void within the $\Lx$--$a$-distribution presented in \citet{Monsch+2019}, within the intrinsic limitations of our 1D approach, which does not yet treat planet formation self-consistently in combination with disc evolution.

This paper is structured as follows. Sect.~\ref{sec:methods} describes the X-ray photoevaporation prescription, as well as the initial setup used for the 1D planet population synthesis model. The outcome of the models is presented in Sect.~\ref{sec:results} and discussed in the context of observational data in Sect.~\ref{sec:discussion}. Finally we draw our conclusion in Sect.~\ref{sec:conclusion}.
\section{Numerical methods}
\label{sec:methods}

\subsection{Photoevaporation model}
\label{sec:PE_model}

\begin{figure*}[t!]
    \centering
    \includegraphics[width=\linewidth]{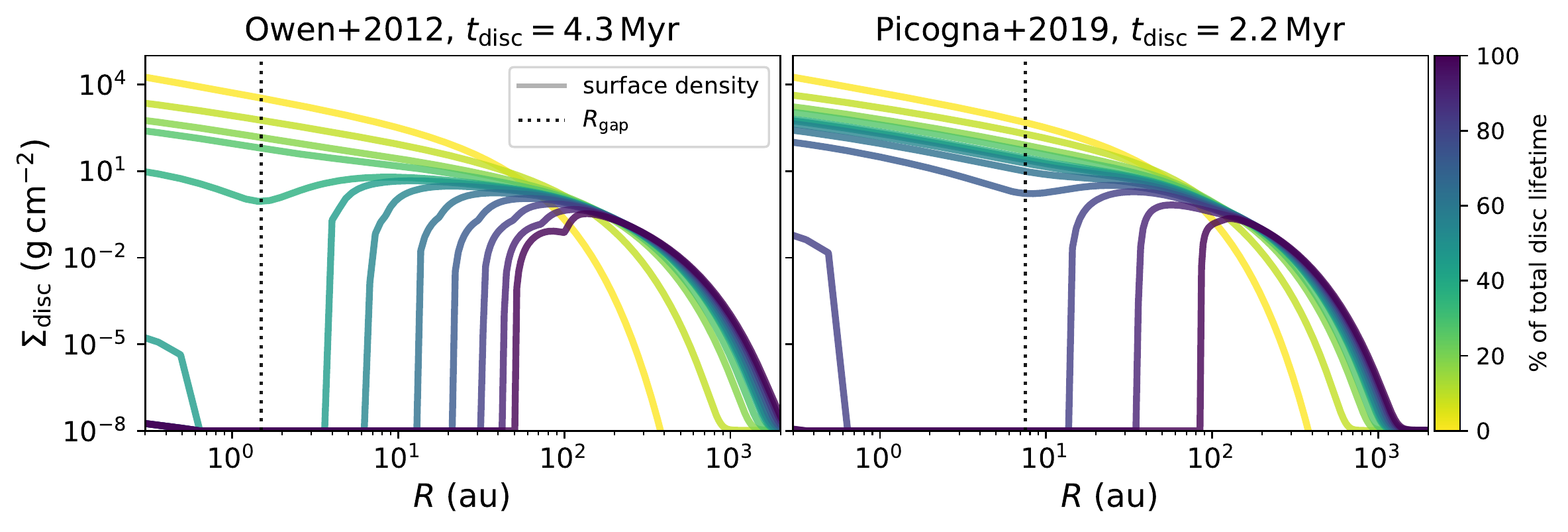}
    \caption{Comparison of the 1D surface density evolution as a function of disc radius for a planet-less disc of $\Mdisc=0.07\,\Msun$, using $\Lx=1\times10^{30}\,\ergs$ for the photoevaporation profiles by \citetalias{Owen+2012} and \citetalias{Picogna+2019}. The different lines are drawn at [0, 25, 50, 60, 70, 72, 75, 80, 85, 90, 95, 99]\,\% of the corresponding total disc lifetime $t_\mathrm{disc}$. The dotted line shows the approximate location of gap opening due to photoevaporation for each model.}
    \label{fig:comp_Sigma}
\end{figure*}

In this study we investigate the impact of XPE on the orbital distribution of a population of giant planets. For this purpose, we first compare the commonly used XPE profiles derived by \citet[][hereafter \citetalias{Owen+2012}]{Owen+2010, Owen+2011, Owen+2012} with the updated one by \citet[][hereafter \citetalias{Picogna+2019}]{Picogna+2019}. 
Both models were computed using radiation-hydrodynamical calculations; however \citetalias{Picogna+2019} include a parameterisation of the temperature as a function of the local gas properties, as well as the column density to the star. Wind mass loss rates ($\Mdotwind$) obtained by \citetalias{Picogna+2019} are approximately a factor of two higher compared to \citetalias{Owen+2012}, and present significant differences in the radial mass loss profile ($\Sigmadotwind$). 
The relevant equations are given in 
Appendix~\ref{sec:AppendixA}, in which Fig.~\ref{fig:comp_PEprofiles} shows a direct comparison of $\dot{M}_\mathrm{w}(\Lx)$ and $\dot{\Sigma}_\mathrm{w}(R)$ for both profiles. 
The one by \citetalias{Picogna+2019} produces not only more vigorous winds (i.e. higher $\dot{M}_\mathrm{w}$), but it also extends further out into the disc (up to $\sim 120\,\au$), therefore leading to the more efficient removal of the disc on a shorter timescale. 
\citet{Woelfer+2019} present similar XPE models, however for low-metallicity discs, which are depleted in carbon. They predict significantly higher gas temperatures and photoevaporative winds in such discs due to the larger penetration depth of the X-rays. 
Investigating the impact of photoevaporation in carbon-depleted discs on the migration process of giant planets is beyond the scope of this paper, but will be attempted in future work.

Fig.~\ref{fig:comp_Sigma} shows the 1D surface density evolution of a planet-less disc with an initial mass of $\Mdisc=0.07\,\Msun$, applying the two different photoevaporation profiles with a reference X-ray luminosity of $\Lx=1\times10^{30}\,\ergs$ for a $0.7\,\Msun$ star.
The total disc lifetimes differ by almost $2\,\Myr$ (i.e. $t_\mathrm{disc,\,Owen}=4.3\,\Myr$ versus $t_\mathrm{disc,\,Picogna}=2.2\,\Myr$), as expected from the enhanced efficiency of the updated photoevaporation profile by \citetalias{Picogna+2019}.
Once the viscous accretion rate onto the star falls below the photoevaporative wind mass loss rate, photoevaporation opens a gap in the disc, cutting the inner disc off from further mass-supply by the outer disc. 
While the profile by \citetalias{Owen+2012} opens a gap at 1--$2\,\au$ between 70--$72\,\%$ of the corresponding total disc lifetime, the profile by \citetalias{Picogna+2019} opens it at slightly larger radii of around 7--$8\,\au$ after 85--$90\,\% \, \times t_\mathrm{disc,\,Picogna}$. 
The latter photoevaporation profile will therefore leave the surface density structure of the disc relatively unperturbed for a larger fraction of the total disc lifetime and cause gap opening at later stages of the disc's global evolution.
The reason for this is that the $\Sigmadotwind$-profile from \citetalias{Owen+2012} peaks around 1--$2\,\au$ and declines relatively steeply beyond that, so that the mass loss will be mostly concentrated near the peak of $\Sigmadotwind$.
Even though the profile by \citetalias{Picogna+2019} peaks even closer inside, it is, in contrast, flatter at larger disc radii, leading to the more efficient removal of material also outside of the peak of $\Sigmadotwind$.
Therefore, the gap will open at later stages compared to the total disc lifetime (which is, nevertheless, significantly shorter for the \citetalias{Picogna+2019} model) and at larger radii in this case. 

From this point on, the disc enters the transition disc phase. As the inner disc is viscously accreted onto the host star, its opacity is reduced quickly (approximately on the viscous timescale), so that the outer disc can be directly irradiated by the central star \citep{Alexander+2006a, Alexander+2006b}. This leads to the very efficient dispersal of the outer disc, roughly on a timescale of a few $10^5\,\yr$.
Both \citetalias{Owen+2012} and \citetalias{Picogna+2019} present different mass loss profiles for primordial and transition discs, which are implemented in our model (Eq.~\ref{eq:Owen_Sigmadotwind_TD} and Eq.~\ref{eq:Picogna_Sigmadotwind_TD}, respectively). 
The switch between both profiles is performed once a gap has opened in the disc and the radial column density inside this gap becomes less than the maximum X-ray penetration depth of $\sim 2.5\times10^{22}\,\mathrm{cm}^{-2}$ \citep{Ercolano+2009a, Picogna+2019}.

\subsection{1D planet population synthesis}
\label{sec:1dpopsynth}

To model the orbital evolution of giant planets in a population of young disc-bearing stars, we used the 1D viscous evolution code \spock. We follow a similar setup as described by \citet{ER15, Jennings+2018} and \citet{Monsch+2021}, which are mostly based on previous models by \citet{Armitage2007}, \citet{AA09}, and \citet{AP12}. We will therefore only briefly summarise our numerical model and refer the reader to \citet{ER15} for a more detailed description of the employed code.

Each model follows the combined evolution of a single giant planet embedded in a protoplanetary disc subject to viscosity and XPE driven by the host star. 
The surface density evolution of the coupled planet-disc system can be described via the equation:

\begin{equation} 
\frac{\partial \Sigma}{\partial t} = \frac{1}{R}\frac{\partial}{\partial R}\left[ 3R^{1/2} \frac{\partial}{\partial R}\left(\nu \Sigma R^{1/2}\right) - \frac{2 \Lambda \Sigma R^{3/2}}{(G M_\star)^{1/2}}\right] - \dot{\Sigma}_{\mathrm {w}}(R,t).
\label{eq:sigma_evol}
\end{equation}
The first term on the right hand side of Eq.~\ref{eq:sigma_evol} describes the viscous evolution of the disc \citep{LyndenBellPringle1974}, the second term treats the migration of the planet \citep{LinPapaloizou1979, LinPapaloizou1986, Armitage+2002} and $\Sigmadotwind(R,t)$ corresponds to the surface mass loss profile due to photoevaporation. Here, $\Sigma(R,t)$ describes the gas surface density of the disc, $M_\star=0.7\,\Msun$ is the stellar mass, $R$ the distance from the star and $G$ is the gravitational constant. 

\begin{figure}
    \centering
    \includegraphics[width=\linewidth]{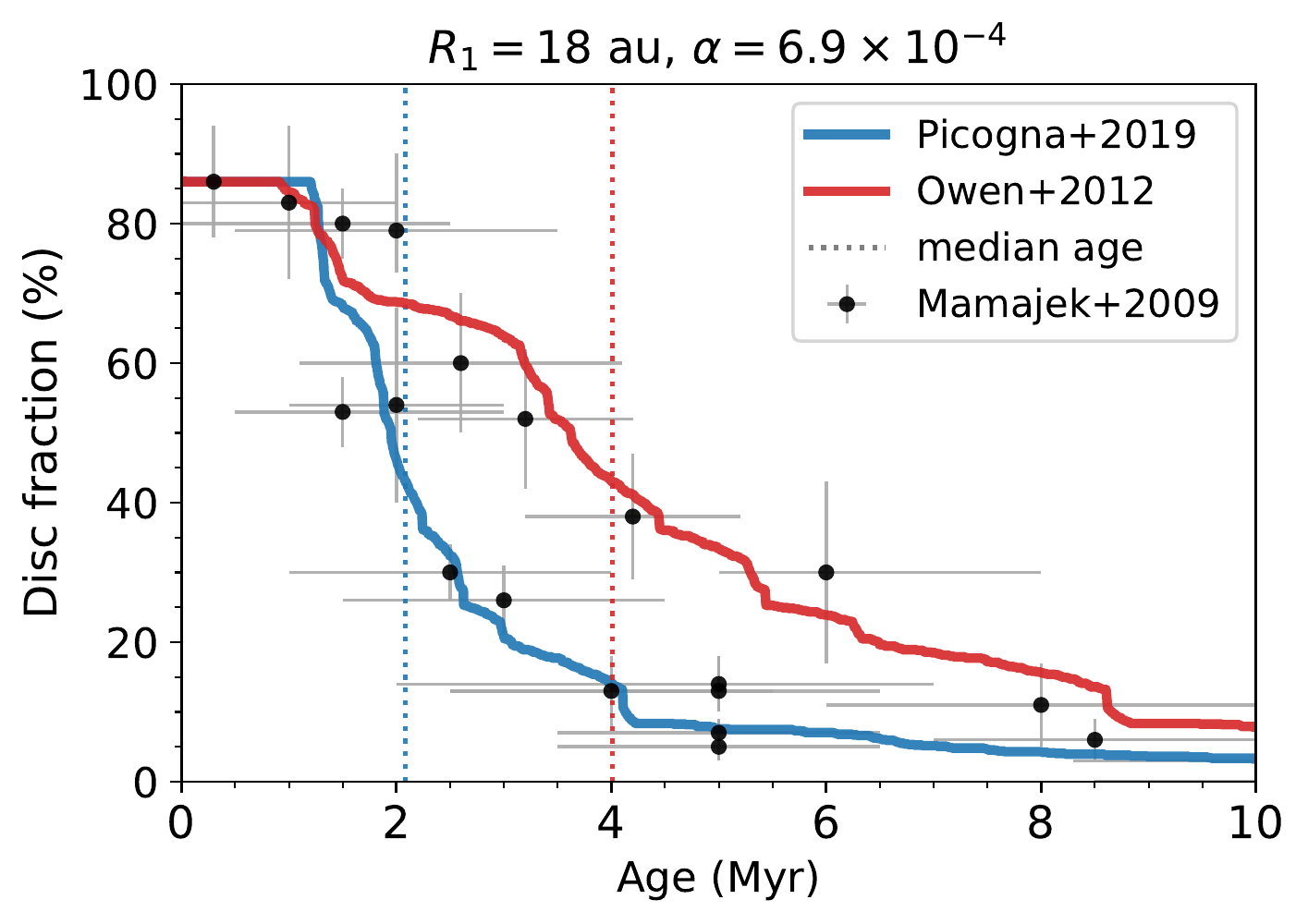}
    \caption{Disc fraction as a function of time from two evolving disc populations using the XPE profiles by \citetalias{Picogna+2019} (solid blue line) and \citetalias{Owen+2012} (solid red line). For each model, 1000 simulations were performed, using different X-ray luminosities that were sampled randomly from the XLF of Taurus. The dotted lines show the corresponding median disc lifetimes of each distribution. The black dots show observed disc fractions compiled by \citet{Mamajek+2009}. The simulated disc fractions were scaled to $86\,\%$ in order to account for binary interactions \citep[cf.][]{Owen+2011}.}
    \label{fig:R1_alpha}
\end{figure}

Eq.~\ref{eq:sigma_evol} was discretised on a grid of 1000 radial cells (which is increased to 4000 at the time of planet insertion), equispaced in $R^{1/2}$ and extending from $0.04\,\au$ to $10^4\,\au$. Further we adopted the self-similarity solution of the diffusion equation by \citet{LyndenBellPringle1974} for the surface density profile of the disc, assuming a time-independent power-law scaling of the disc radius with the kinematic viscosity, $\nu\propto R^\gamma$, where $\gamma=1$ \citep[cf. Sect.~4 in][]{Hartmann+1998}:

\begin{equation}
\label{eq:surfdens}
    \Sigma(R, t=0) = \frac{\Mdisc(t=0)}{2\pi R_\mathrm{1} R}\,\exp\left( -\frac{R}{R_\mathrm{1}} \right).
\end{equation}
Here, $\Mdisc(t=0)=0.07\,\Msun$ is the initial disc mass at time zero. The disc scaling radius $R_\mathrm{1}$ describes the location of the exponential cutoff of the surface density profile. Together with the kinematic viscosity $\nu=\alpha c_\mathrm{s}H$, where $c_\mathrm{s}$ is the sound speed of the gas, $H$ the disc scale height, and $\alpha$ the dimensionless Shakura-Sunyaev parameter \citep{ShakuraSunyaev1973}, it sets the viscous timescale $\tnu=R_\mathrm{1}^2/(3\nu)$. We assume locally isothermal discs with an aspect ratio of $H/R=0.1$ at $R_1$, which results in flared discs following $H \propto R^{5/4}$ and a midplane temperature structure scaling as $T_\mathrm{mid}\propto R^{-1/2}$, so that $T_\mathrm{mid}\approx[2100\,\mathrm{K}, 4\,\mathrm{K}]$ at the inner and outer boundary, respectively.

The values for $R_1$ and $\alpha$ need to be chosen such that, combined with the effect of viscous accretion and mass loss due to photoevaporation, observationally supported median disc lifetimes ranging between 1--$3\,\Myr$ are obtained \citep[e.g.][]{Haisch+2001, Mamajek+2009, Fedele+2010, Ribas+2014, Ribas+2015}.
We therefore followed the approach described by \citet{Owen+2011} and constructed populations of evolving protoplanetary discs using different combinations of $R_1$ and $\alpha$. 
The results from this test are summarised in Fig.~\ref{fig:R1_alpha}, which shows the disc fraction as a function of time from both XPE models tested in our study, using in total 1000 individual simulations in which the X-ray luminosities were sampled stochastically from the X-ray luminosity function (XLF) of the Taurus cluster \citep{Guedel+2007}. Following \citet{Owen+2011}, the resulting distributions were scaled to $86\,\%$, assuming an initial close binary fraction of $14\,\%$ for NGC~2024 \citep{Haisch+2001}. As it has an estimated age of $0.3\,\Myr$, this stellar cluster is likely too young for discs to have been destroyed by planet formation or photoevaporation entirely. 
We found that $R_\mathrm{1}=18\,\au$ and $\alpha=6.9\times10^{-4}$, which yields a viscous timescale of $\tnu=7\times10^{5}\,\yr$ at $R_\mathrm{1}$, reproduces the observed disc fractions compiled by \citet{Mamajek+2009} best. Due to the increased photoevaporative mass loss rates, the profile by \citetalias{Picogna+2019} generates shorter median disc lifetimes than the one by \citetalias{Owen+2012}; however, both lie well within the observed spread in the disc fractions. In order to extract the effect of the XPE profile itself on the resulting orbital distribution of giant planets, we kept the same initial disc profile for all simulations in the remainder of this paper, regardless of the photoevaporation profile.

While we employ a value of disc viscosity, which is roughly consistent with recent observations of low disc turbulence \citep[e.g.][]{Flaherty+2018}, it is important to notice that realistic disc lifetimes could also be achieved by using different combinations of $R_1$ and $\alpha$ within one model as there is no a priori reason for them to be fixed in a population of discs. This approach was for example followed in a study similar to ours performed by \citet{Ercolano+2018}, who found, however, that their results do not change qualitatively, showing the robustness of these models against the specific choice of the underlying disc viscosity.
Nevertheless, the implications of using a higher value of disc viscosity, as was for example done by \citet{AP12}, are explored in detail in Appendix~\ref{sec:viscosity}.

We modelled giant planets with masses ranging from 0.5--$5\,\Mjup$, which were inserted between 5--$20\,\au$ into the disc. The choice for the range of insertion locations is solely based on the simple assumption that most giant planets form outside the water snow-line due to more favourable initial conditions \citep[e.g.][]{KennedyKenyon2008, Guilera+2020}. Around solar analogues, the water snow line is expected to lie between 2--$5\,\au$ \citep{Mulders+2015b}, therefore assuming $5\,\au$ as the minimum planet `formation' location is a rather conservative choice.
While the planets are allowed to accrete mass from the disc, their formation itself is not simulated in our code. Therefore, the formation times of the planets were drawn randomly from a uniform distribution between $0.25~\Myr$ \citep[which we assume to be the minimum time required to form a gas giant in the core accretion paradigm, cf.][]{Pollack+1996} and $\tclear$, where

\begin{equation}
    \tclear = \frac{t_\mathrm{\nu}}{3} \left( \frac{3 M_\mathrm{d}(t=0)}{2t_\mathrm{\nu} \dot{M}_\mathrm{w}} \right)^{2/3},
    \label{eq:tclear}
\end{equation}
is the time at which photoevaporation starts to clear the disc \citep{Clarke+2001, Ruden2004}.
For $\Lx=1\times10^{30}\,\ergs$, this corresponds to a disc clearing timescale of approximately $\tclear=1.8\,\Myr$ for the profile by \citetalias{Owen+2012} and $\tclear=1\,\Myr$ for the profile by \citetalias{Picogna+2019}. We further ensure that all discs are massive enough to actually form a giant planet via core accretion (if the planet is, for example, inserted at late stages of disc evolution), and therefore require the dust disc mass to be $\Sigma_\mathrm{dust}=0.01\Sigma_\mathrm{gas}\geq 10\,M_\oplus$ at the time of planet insertion.

Planet accretion is modelled following Eq.~5 in \citet{VerasArmitage2004}, whose implications will be discussed in more detail in Appendix~\ref{sec:planet_accretion}.
The planets then migrate in the disc following the impulse approximation \citep{LinPapaloizou1979, LinPapaloizou1986, Armitage+2002}:

\begin{equation}
\frac{\mathrm{d}a}{\mathrm{d}t} = -\left(\frac{a}{G M_\star} \right )^{1/2} \left(\frac{4\pi}{M_\mathrm{p}}\right) \int_{R_\mathrm{in}}^{R_\mathrm{out}}{R\Lambda\Sigma}\mathrm{d}R, \mathrm{~where}
\label{eq:impulse_approx}
\end{equation}

\begin{equation}
\Lambda(R,a) = \left\{ \begin{array}{ll}
- \frac{q^2 G M_\star}{2R} \left(\frac{R}{\Delta_{\mathrm p}}\right)^4 & \textrm{if } \, R < a\\
\frac{q^2 G M_\star}{2R} \left(\frac{a}{\Delta_{\mathrm p}}\right)^4 & \textrm{if } \,R > a.\\
\end{array}\right. 
\label{eq:impulse_torques}
\end{equation}
Eq.~\ref{eq:impulse_approx} describes the evolution of the planetary semi-major axes as a function of time, the underlying 1D disc surface density, $\Sigma$, and the rate of the specific angular momentum transfer from the planet to the disc (i.e. the specific torques), $\Lambda (R, a)$. 
Here, $q=\Mpl/M_\star$ is the planet to star mass-ratio and $\Delta_\mathrm{p} = \max(H,|R-a|)$ is the impact parameter, which ensures that only material outside of one disc scale height, $H=0.1R$, is included into the torque calculation.
We implemented the proposed fix by \citet{Monsch+2021} to the impulse approximation, which parks the planet as soon as the maximum surface density inside the planet location becomes $\Sigma \leq 10^{-6}\,\gcm$ (i.e. the inner disc is dispersed) and $\Sigma \leq 10^{-2}\,\gcm$ at the 3:2 resonance location outside the planet (to make sure that the outer disc is depleted enough to not continue pushing the planet inside). 
Further, each simulation is at the latest stopped at $t=10\,\Myr$ or once the planet reaches $a\leq0.15\,\au$ as we do not attempt to model the formation of hot Jupiter systems. 

\begin{figure}[t!]
    \centering
    \includegraphics[width=\linewidth]{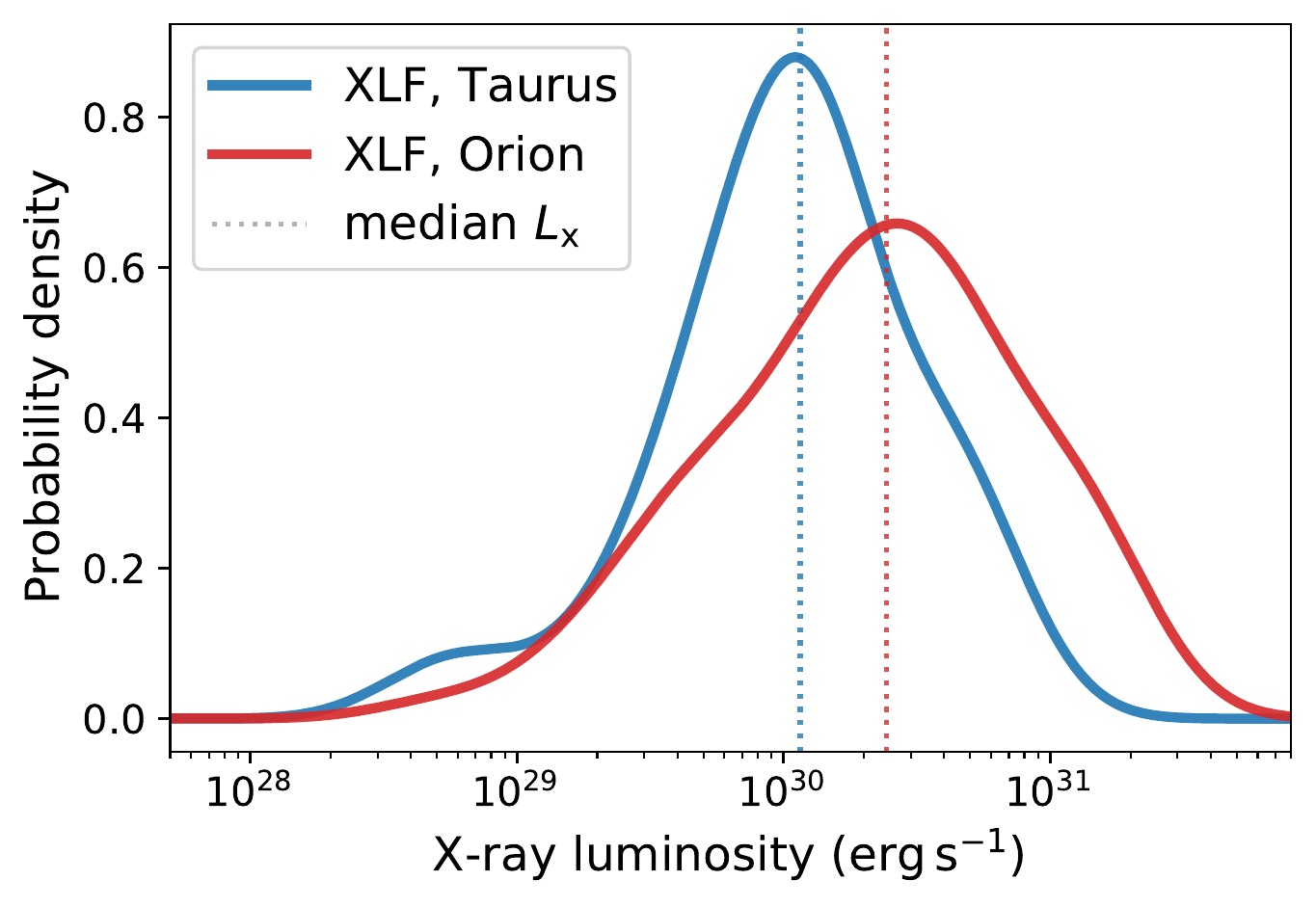}
    \caption{XLFs for pre-main-sequence stars located in the Taurus Cluster \citep[$0.5$--$1\,\Msun$,][]{Guedel+2007, Owen+2011} and the Orion Nebula Cluster \citep[$0.5$--$0.9\,\Msun$,][]{PreibischFeigelson2005}. The dotted lines are drawn at the median X-ray luminosity for each corresponding XLF.}
    \label{fig:Lx_func}
\end{figure}

We considered observationally motivated X-ray luminosities following XLFs for pre-main-sequence stars in the Taurus cluster \citep[$M_\star=0.5$--$1\,\Msun$,][]{Guedel+2007, Owen+2011} and the Orion Nebula Cluster \citep[$M_\star=0.5$--$0.9\,\Msun$,][]{PreibischFeigelson2005}, which are shown in Fig.~\ref{fig:Lx_func}. 
The difference between both XLFs is relatively small and lies mostly towards higher X-ray luminosities, which can be related to the different treatment of stellar flares \citep[see][for a detailed discussion]{Owen+2011}.
Both cover a spread of about two orders of magnitude in X-ray luminosities, but in order to study the full extent in $\Lx$-parameter space in our simulations, we sampled the X-ray luminosities linearly between $\log(\Lx/\ergs)=27$--32 (i.e. both extreme ends of the XLFs) using in total 1000 bins. To facilitate the identification of any $\Lx$-specific features, we further oversampled given $\Lx$-ranges (depending on the simulation) with another 500 bins. 
Table~\ref{table:sim_initcond} summarises the initial conditions for the discs modelled in our study, while 
Table~\ref{tab:sim_summary} collects the setups used for the different simulations.

\begin{table}[]
\centering                          
\begin{tabular}{l|l}        
\hline\hline                 
Parameter & Value\\
\hline                        
$M_\star~(\Msun)$ & $0.7$ \\
$\Mdisc~(\Msun)$ & $0.07$ \\
$\alpha$ & $6.9\times 10^{-4}$ \\
$R_\mathrm{s}$ (au) & $18.$ \\
$H/R$ & 0.1 \\
$\Mpl~(\Mjup)$ & [$0.5,5.$] \\
$t_\mathrm{form}$ (Myr) & [$0.25,\tclear$]\\
$R_\mathrm{in}$ (au) & 0.04 \\ 
$R_\mathrm{out}$ (au) & $10^4$ \\  
$\log(\Lx/\ergs)$ & [$27,32$] \\
\hline                                   
\end{tabular}
\caption{Initial conditions for the \spock\ simulations described in Sect.~\ref{sec:1dpopsynth}.}
\label{table:sim_initcond}
\end{table}

\begin{table*}[ht!]
    \centering
    \begin{tabular}{c|ccccc}
    \hline\hline
        Name & $a_\mathrm{insert}$ & PE-profile & oversampled $\log\Lx$ & $a_\mathrm{final} = 0.15\,\au$ & comments\\
        \hline 
        \texttt{Owen\_5au} & $5\,$au & \citetalias{Owen+2012} & 28.7--31.0 & 33.9\,\% &\\
        \texttt{Owen\_10au} & $10\,$au & \citetalias{Owen+2012} & 28.7--31.0 & 23.1\,\% &\\
        \texttt{Owen\_20au} & $20\,$au & \citetalias{Owen+2012} & 28.0--29.6 & 10.1\,\% &\\
        \texttt{Owen\_IPMF} & 5--$20\,$au & \citetalias{Owen+2012} & 29.0--31.0 & 17.8\,\% & uniform-random sampling of $a_\mathrm{insert}$ \& \\ &&&&& planet mass sampled from IPMF\\
        \texttt{Picogna\_5au} & $5\,$au & \citetalias{Picogna+2019} & 28.0--30.5 & 40.3\,\% &\\
        \texttt{Picogna\_10au} & $10\,$au & \citetalias{Picogna+2019} & 28.0--30.5 & 25.1\,\% &\\
        \texttt{Picogna\_20au} & $20\,$au & \citetalias{Picogna+2019} & 28.0--29.5 & 12.6\,\% &\\
        \texttt{Picogna\_IPMF} & 5--$20\,$au & \citetalias{Picogna+2019} & 28.0--29.5 & 23.0\,\% &  uniform-random sampling of $a_\mathrm{insert}$ \& \\ &&&&& planet mass sampled from IPMF\\
        \hline
    \end{tabular}
    \caption{Summary of the setups used for the different population synthesis models presented in Sect.~\ref{sec:results}. }
    \label{tab:sim_summary}
\end{table*}
\begin{figure*}[ht!]
\centering
\includegraphics[width=0.9\linewidth]{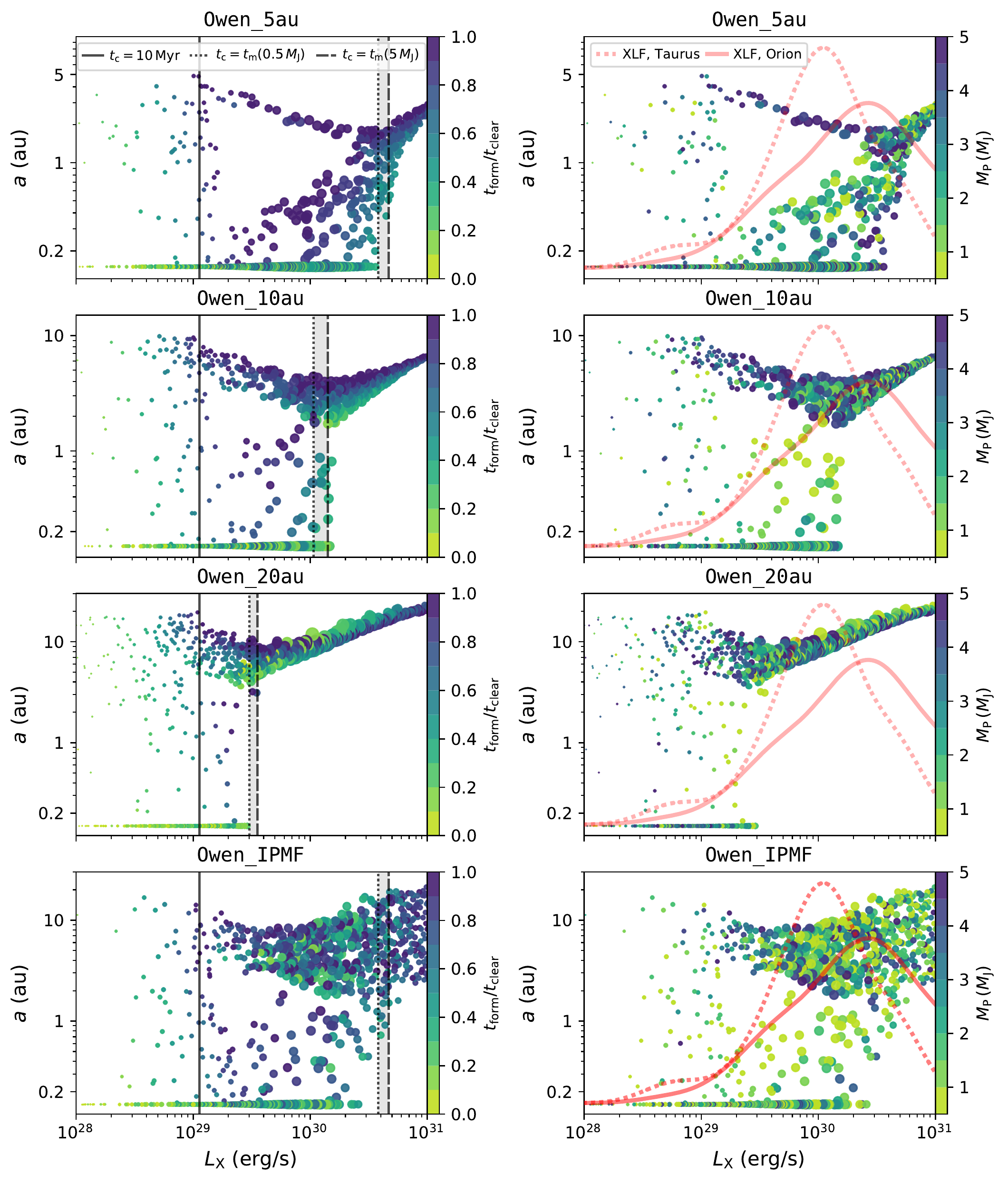}
\caption{Comparison of the final $\Lx$--$a$-distributions for the \texttt{Owen\_XX} models using the photoevaporation profile by \citetalias{Owen+2012}. The different rows show the outcomes of the population synthesis models using different insertion locations of the planets ($5\,\au$, $10\,\au$, $20\,\au$ and random insertion locations). Colours in the left column correspond to the formation time of the planet with respect to the disc clearing time due to photoevaporation. In the right column, colours reflect the initial planetary mass. The black lines highlight the different regimes, in which the final planet parking location is set by different effects. These are discussed in detail in Sect.~\ref{sec:popsynth_Owen2012_5au}. 
The red lines in the right panels correspond to the XLFs as shown in Fig.~\ref{fig:Lx_func}. Additionally, the data points were weighted linearly following the XLF in Taurus so that their size reflects the observing probability at a given $\Lx$. This step was performed in order to emphasise which regions in $\Lx$-parameter space are strongly over-crowded due to the linear sampling of $\log\Lx$ in our model.} 
 \label{fig:popsynth_Owen2012}
\end{figure*}

\section{Results}
\label{sec:results}

Figures~\ref{fig:popsynth_Owen2012} and \ref{fig:popsynth_Picogna2019} collect the outcome from the different population syntheses, each performed with in total 1500 disc-planet systems employing the photoevaporation profiles by \citetalias{Owen+2012} and \citetalias{Picogna+2019}. 
Each row shows the resulting $\Lx$--$a$-distribution for different insertion locations of the planets (ranging from 5--$20\,\au$; see Table~\ref{tab:sim_summary}). 
The colours in the left column reflect the formation time of the planets relative to the disc clearing time (given by Eq.~\ref{eq:tclear}), so that $t_\mathrm{form}/\tclear < 1$, while in the right column colours correspond to the initial planetary mass. 
Further, the XLFs shown in Fig.~\ref{fig:Lx_func} were over-plotted in the right panels to demonstrate which parts of $\Lx$-parameter space have been oversampled strongly due to the linear sampling of $\log \Lx$ in our model. 
The sizes of the data points additionally scale linearly with the value of the XLF of Taurus at the corresponding $\Lx$ to emphasise which part of $\Lx$-parameter space would be most likely to be observed in a true sample. 
The crowding of planets at $0.15\,\au$ is an artefact resulting from our numerical setup, at which the simulations are forced to stop. In reality, however, these planets would either end up as hot Jupiters or be engulfed by their host star. The fraction of planets that reached the inner grid boundary are summarised for each simulation in Table~\ref{tab:sim_summary}.
The following subsections will discuss each row of Fig.~\ref{fig:popsynth_Owen2012} and Fig.~\ref{fig:popsynth_Picogna2019} separately.

\subsection{Model \texttt{Owen\_5au}}
\label{sec:popsynth_Owen2012_5au}

For the population synthesis presented in the top row of Fig.~\ref{fig:popsynth_Owen2012}, planets were inserted at $5\,\au$ into the disc. 
The black lines highlight three different regimes in the $\Lx$--$a$-distribution, in which the final location of the giant planets is dominated by different effects:

\begin{enumerate}
    \item $\tclear \gg 10\,\Myr$ (left)
    \item $\tclear \ll \tmigr$ (right)
    \item $10\,\Myr > \tclear > \tmigr$ (centre)
\end{enumerate} 
These will be discussed in detail in the following.

\subsubsection{$\tclear \gg 10\,\Myr$ (left)}

For $\Lx \lesssim 10^{29}\,\ergs$, the disc clearing timescale becomes larger than $10\,\Myr$, which is the time at which our simulations are forced to stop (e.g. $\tclear\approx63\,\Myr$ for $\Lx=10^{28}\,\ergs$). 
In this regime, the type~II migration timescale of the planets is much shorter than the disc clearing time, meaning that the surface density has already decreased significantly due to viscous accretion onto the host star once photoevaporation sets in and starts clearing the disc. 
At this point, the ratio of the disc mass to planet mass is so low that the planets have already stopped migrating, before photoevaporation could possibly affect or even halt their inward migration. Consequently, for low $\Lx$, photoevaporation becomes ineffective in parking giant planets in the disc, and therefore they simply continue migrating until the simulations are forced to end at $10\,\Myr$.
Thus, for $\Lx \lesssim 10^{29}\,\ergs$, planets randomly populate semi-major axes between $5\,\au$ and the star, depending on their formation time and initial mass. In reality, however, planets in very weakly photoevaporating discs would continue to form and migrate inwards until accretion onto the star causes the surface density to become low enough to halt planet formation and migration. Thus, while the random population in our model is a direct consequence of the maximum disc lifetime assumed in our simulations, the distribution of planets in this regime is still expected to be random if planet formation were to be treated self-consistently within our model.

\subsubsection{$\tclear \ll \tmigr$ (right)}

For $\Lx \gtrsim 5\times10^{30}\,\ergs$, $\tclear$ becomes shorter than the migration timescale of the most massive planets in our numerical model, meaning that photoevaporation produces such vigorous winds in this regime that it disperses the discs, before the planets could cross $\Rgap$. Further, the range for the possible formation times of the planets becomes very small for high $\Lx$ as $\tclear$ is only marginally larger than $0.25\,\Myr$ (e.g. $\tclear = 0.33\,\Myr$ for $\Lx=10^{31}\,\ergs$), which is the minimum time required to form a giant planet in our model. Thus, for increasing X-ray luminosities, photoevaporation becomes more efficient in dispersing the circumstellar material and consequently parking the planets quickly after they are inserted into the disc, creating a diagonally shaped tail towards higher $\Lx$.

\subsubsection{$10\,\Myr > \tclear > \tmigr$  (centre)}

In this regime, the disc clearing timescale is shorter than $10\,\Myr$, meaning that for each X-ray luminosity within this range, disc dispersal via XPE will be initiated before the simulations reach their maximum run time. 
Further, the migration timescale, $\tmigr$, for a planet of $0.5\,\Mjup$ (dotted line) or $5\,\Mjup$ (dashed line) that is formed at the earliest possible time of $0.25\,\Myr$ in our model (assuming a disc without photoevaporation), becomes shorter than the disc clearing time. This means that the planet may reach the inner boundary (depending on its insertion time and mass) before photoevaporation starts clearing the disc. 
Therefore, the X-ray luminosities ranging between $\tclear=\tmigr(0.5\,\Mjup)$ and $\tclear=\tmigr(5\,\Mjup)$ can be considered as an upper limit, for which a planet of given mass in our setup is potentially able to cross $\Rgap$, before photoevaporation could potentially open a gap at this location. 
In contrast, for higher $\Lx$, photoevaporation will disperse the disc before planets can cross $\Rgap$, so that all planets are parked soon after they have been inserted into the disc. 
Therefore, $10\,\Myr > \tclear > \tmigr(5\,\Mjup)$ defines the range in $\Lx$-parameter space, in which one would expect to observe an under-density of planets in the $\Lx$--$a$-distribution, caused by disc dispersal via photoevaporation. This is because photoevaporation opens an annular gap at $\Rgap$ in the disc on timescales comparable to the migration timescales of the planets, which will force the planets to either edge of the gap, with the majority sneaking by before gap opening. This creates a void of planets in the observed $\Lx$--$a$-distribution, which is centred on $\Rgap$.

Indeed, for intermediate X-ray luminosities of $\sim 10^{29}\,\ergs$ to $\sim 5\times10^{30}\,\ergs$, a large triangular-shaped desert of planets centred on $\sim 1\,\au$ can be observed. 
The insertion location of planets lies close to $\Rgap$ in the \texttt{Owen\_5au} model, so that only planets that formed late relative to the disc clearing time (i.e. $t_\mathrm{form}/\tclear \gtrsim 0.7$) are parked outside of the gap. 
Assuming that the planet formation efficiency at $5\,\au$ right before the onset of disc dispersal is considerably low, such planets could for example correspond to planets that formed earlier in the outermost parts of the planet-forming disc, possibly due to gravitational instability or pebble accretion, and have migrated up to $5\,\au$ just before photoevaporative disc clearing was initiated.
It is therefore highly suggestive that the insertion location of the planets plays the most important role in determining the final parking location of the planets in our model. To investigate this further, we will later discuss the impact of different planet insertion locations on our results.

\subsubsection{Planet-mass distributions}

As inferred from Eq.~\ref{eq:impulse_approx}, the migration rate of a planet depends on its mass. Therefore it is to be expected that distinct trends for planets of different masses can be observed in the orbital distribution of giant planets as more massive planets will reach higher migration rates, therefore reaching smaller radii before photoevaporation starts clearing the disc. 

From the top right panel in Fig.~\ref{fig:popsynth_Owen2012} it becomes apparent that planets with $\Mpl \gtrsim 2.5\,\Mjup$ accumulate outside of the observed void, which is roughly centred on $1\,\au$. These planets are so massive that they strongly suppress the inflow of material across the planetary orbit, which reduces the opacity of the inner disc such that photoevaporation can start clearing the inner disc earlier as would be expected without the presence of a giant planet. This effect is termed Planet-Induced PhotoEvaporation (PIPE) and was first identified by \citet{AA09} and \citet{Rosotti+2013} as a direct consequence of the strong coupling between planet formation and protoplanetary disc clearing. 
As photoevaporation opens a gap in the disc, it fully decouples the inner from the outer disc, so that any further inflow of material from the outer disc is inhibited. For planets triggering PIPE, gap opening due to photoevaporation is therefore always initiated before they can cross $\Rgap$, leading to their pile-up just outside of this location. In contrast, all lower-mass gas giants with $\Mpl \lesssim 2.5\,\Mjup$ are able to cross this location before gap opening as they do not trigger PIPE. Therefore they will either pile up just inside the gas-free cavity or keep migrating inside, depending on how fast the inner disc is dispersed. 
The final parking location of planets located inside of $\Rgap$ at the time of gap opening therefore ultimately depends on their formation time, meaning only planets that are formed late and consequently also crossed $\Rgap$ at later stages, will be able to survive, while all remaining planets migrate all the way onto the star and therefore pile up at $0.15\,\au$ in our model. 
This result is consistent with recent work by \citet{Emsenhuber+2020}, who also found that giant planets need to acquire their full mass up until shortly before the dispersal of the disc to prevent their strong inward migration, which would otherwise bring them to the inner edge of the disc.

\subsection{Model \texttt{Owen\_10au}}
\label{sec:popsynth_Owen_10au}

The numerical setup used to obtain the results shown in the second row of Fig.~\ref{fig:popsynth_Owen2012} is conceptually the same as for model \texttt{Owen\_5au}; however, now planets were inserted at $10\,\au$ into the disc. 
Even though a void of planets is observed again in the $\Lx$--$a$-distribution, its size is significantly reduced compared to the previous model. This is easily explained because, due to the longer migration timescales of planets embedded at $10\,\au$, photoevaporation will have more time to reduce the disc surface density and therefore eventually park the planets before they can cross $\Rgap$.
In the previous model, only very massive planets that reached the inner disc at late times (i.e. planets that were formed late in our model) accumulated outside of the void. Now planets with a broader range of masses and formation times ($\tform/\tclear \gtrsim 0.5$) can be parked outside of the photoevaporative gap. Consequently also the total number density of planets inside $1\,\au$ is significantly reduced compared to the previous model.

As expected, left of $\tclear=10\,\Myr$ no significant difference in the planet distribution can be observed (except of the higher number density of planets since fewer planets migrate up to $0.15\,\au$) as photoevaporation does not play a significant role in this regime. Just as was observed in the previous model, also here planets are parked at random semi-major axes. However, the right edge of the void, which is confined by $\tclear=\tmigr$, has now significantly moved towards lower X-ray luminosities. This creates an even sharper and longer tail of planets towards higher $\Lx$ as, due to the longer migration timescales, fewer planets will be able to cross $\Rgap$ before XPE starts clearing the disc. The final parking locations will therefore solely be set by the disc clearing timescale, which decreases with higher $\Lx$.

\subsection{Model \texttt{Owen\_20au}}
\label{sec:popsynth_Owen_20au}

In this model, the planets were inserted at $20\,\au$ into the disc. The void that was observed in the previous two models has now shrunken dramatically and only extends from $\sim 1\times10^{30}\,\ergs$ to $4\times10^{30}\,\ergs$.
Right of $\tclear=\tmigr$, the final parking locations of the planets are solely set by the disc clearing time as the migration timescales of planets, which are formed at $20\,\au$, is significantly longer than for planets inserted at $10\,\au$ or $5\,\au$. 
In this regime, photoevaporation has now significantly more time to disperse the disc and open a gap, before any planet could cross this location. Consequently also the mass-distribution of planets outside of the void has become wider as more and more lower-mass planets can be parked outside $\Rgap$, which ultimately results in a reduced number density of planets inside of $1\,\au$.

\subsection{Model \texttt{Owen\_IPMF}}
\label{sec:popsynth_Owen_random}

The previously presented models are extremely idealised. Firstly they assume that planets form at a single location, while in reality, planets are expected to form over a broad range of radii within the planet-forming disc. However, each insertion location of the planet leaves distinct features in the observed $\Lx$--$a$-distribution and it becomes clear that the outcome of our population synthesis models is therefore extremely sensitive to the assumed formation location of the planets.
Secondly, the planet mass was sampled randomly from a flat distribution, ranging from 0.5--$5\,\Mjup$. 
Such an approach is reasonable to ensure that all planet-mass bins contain a statistically significant number of planets, which ultimately enables us to identify any planet-mass related features in the final $\Lx$--$a$ distribution (such as the pile-up of higher-mass planets outside the void in the \texttt{Owen\_5au} model). 
However, the true distribution of giant planets is strongly non-uniform as observational data suggest that it declines with planetary mass, approximately following $\mathrm{d}N/\mathrm{d}\Mpl \propto \Mpl^{-\gamma}$, with $\gamma\approx1$ 
\citep[e.g.][]{Marcy+2005, Ananyeva+2020}.\footnote{We note that this planet mass function was derived from observational data of planet hosts with ages $\sim \Gyr$ and may therefore not represent the primordial mass distribution of exoplanets \citep[see for example][]{Carrera+2018}. However, global planet population synthesis approaches, such as the Bern model, seem to reproduce the observed $1/\Mpl$-planet mass distribution reasonably well \citep[see Fig.~5 in][which is adapted from \citet{Mordasini+2012a}]{Benz+2014}.} 
Our previous approach is therefore not representative of the true sample of giant planets, and our results using a flat initial planet mass function (IPMF) should be treated with caution when directly compared to an observational sample.
Nevertheless, in order to understand how disc dispersal via photoevaporation may affect the migration history of planets, one must first understand how these different initial conditions in our numerical model impact the final outcome. 

Thus, in a new population synthesis we sampled the insertion location of the planets randomly from a uniform distribution between 5--$20\,\au$, and the planet mass from a $1/\Mpl$-distribution as is suggested by observations. 
We emphasise that we do not make any assumptions on how the planets in our model form, but solely assume that most giant planets need to form somewhere outside the water snow line, in order to acquire their gaseous atmosphere.
While this approach is still highly idealised, it is nevertheless useful for understanding what kind of feature would be expected in the observed $\Lx$--$a$-distribution of giant planets. 

It becomes apparent from the lowest left panel of Fig.~\ref{fig:popsynth_Owen2012} that the narrow, diagonally shaped tail, which was present in all previous models has now almost disappeared entirely. 
This is because the \texttt{Owen\_IPMF} model is expected to be a superposition of the previous models with single planet insertion locations. Consequently, for each given insertion location one would expect the tail to shift along the y-axis and further change its length, leading to the random population of the tail between the upper boundary set by the \texttt{Owen\_20au} model, and the lower boundary set by the \texttt{Owen\_5au} model. 
Due to the larger amount of planets now inserted at larger distances from the star, mostly low-mass planets can cross the photoevaporative gap location before gap opening, while the majority of especially higher-mass planets gets parked outside of it. The reason for this is that with increasing insertion location, only planets with decreasing mass can cross the gap location as was seen in the \texttt{Owen\_5au} model. 
Additionally, due to the more realistic planet mass-sampling in this population synthesis, the resulting sample now includes significantly more lower-mass planets, which do not trigger PIPE. Therefore, most of the planets will cross $\Rgap$ before photoevaporation becomes dominant and opens a gap.

\subsection{Model \texttt{Picogna\_5au}}
\label{sec:popsynth_Picogna2019}

\begin{figure*}
\centering
\includegraphics[width=0.9\linewidth]{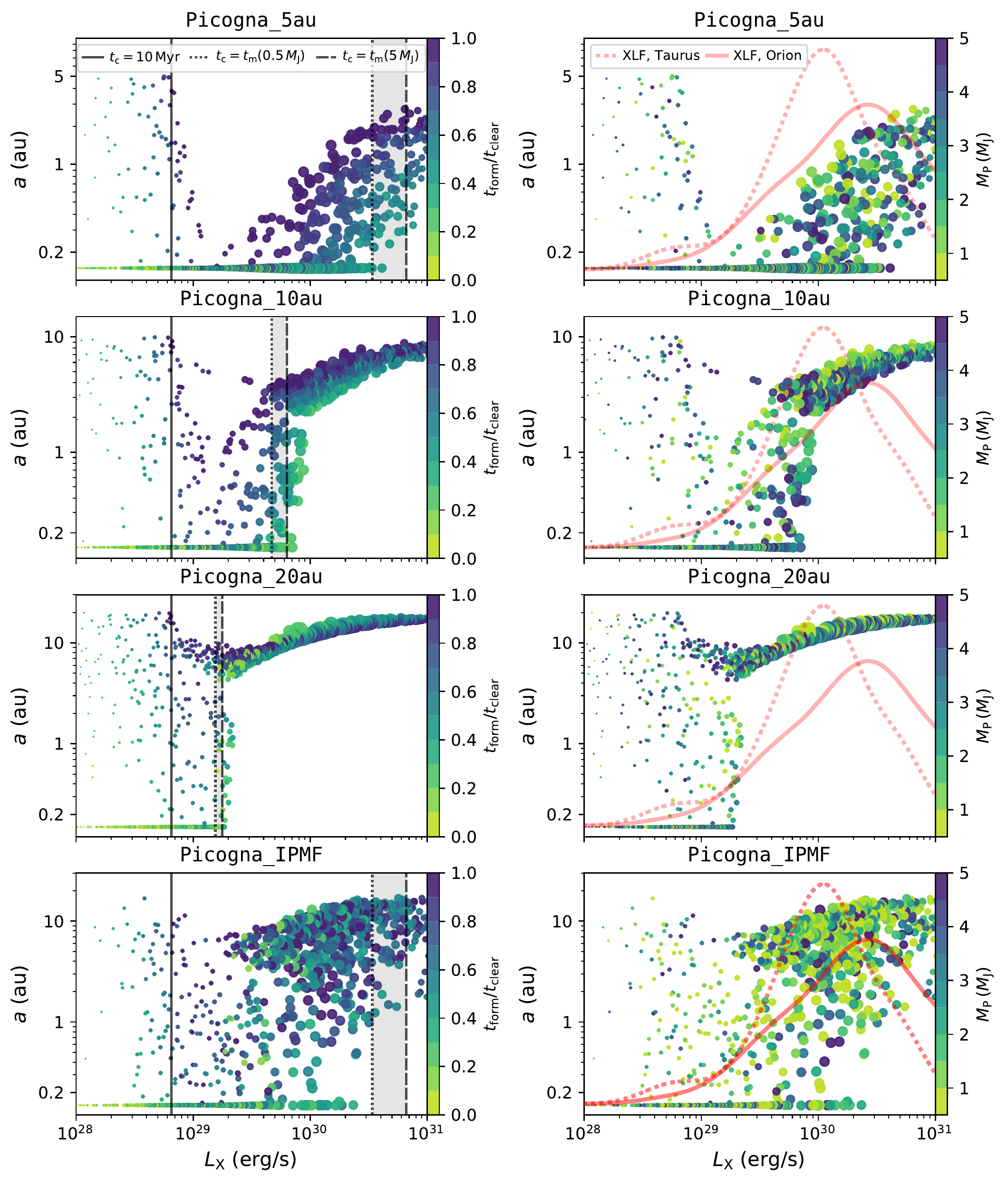}
\caption{Same as Fig.~\ref{fig:popsynth_Owen2012}, but now the photoevaporation profile from \citetalias{Picogna+2019} has been applied.}
 \label{fig:popsynth_Picogna2019}
\end{figure*} 

Fig.~\ref{fig:popsynth_Picogna2019} is conceptually similar to Fig.~\ref{fig:popsynth_Owen2012}; however now the photoevaporation profile by \citetalias{Picogna+2019} was applied. 
A similar triangular void, such as the one identified in the \texttt{Owen\_5au} model, can be observed in the intermediate $\Lx$-regime. However, now the pile-up of massive planets outside of the void is entirely missing. This can be explained by the fact that the photoevaporation profile by \citetalias{Picogna+2019} causes the photoevaporative gap to open at larger radii of 7--$8\,\au$ compared to the profile by \citetalias{Owen+2012}, which opens the gap between 1--$2\,\au$ (cf. Fig.~\ref{fig:comp_Sigma}). This means that in this model, planets are inserted already inside of $\Rgap$, meaning that no planet can be parked outside of it. Consequently, also no tail of planets extending towards higher $\Lx$ can be observed. 
Even though photoevaporation still impacts planet migration by reducing the disc surface density, which in turn affects their migration rates, the final parking locations of the planets in this given model will be mainly set by their formation times as well as their mass. 

The region at which photoevaporation becomes ineffective in affecting planet migration (i.e. left of $\Lx(\tclear=10\,\Myr$)) has now shifted towards a lower $\Lx$ of $\sim 7\times10^{28}\,\ergs$. The reason for this is that the cumulative mass loss rate following from the \citetalias{Picogna+2019} profile is higher than the one from \citetalias{Owen+2012} for such low $\Lx$, therefore strongly increasing the impact of photoevaporation in this regime.
In contrast, the region at which the clearing timescale becomes shorter than the migration timescale of the planets (i.e. right of $\Lx(\tclear=\tmigr$)) has now shifted towards higher $\Lx$; however, also showing a larger range of the migration timescale for the lowest- and highest-mass planets in our model. 
While the profile by \citetalias{Picogna+2019} generally predicts higher mass loss rates, it saturates towards $10^{-7}\,\Msunyr$ for high $\Lx$ (see the left panel of Fig.~\ref{fig:comp_PEprofiles}), while the one by \citetalias{Owen+2012} would predict an exponential increase in $\Mdotwind$ in this regime. This means that for $\Lx \gtrsim  5\times10^{30}\,\ergs$ the impact of photoevaporation on planet migration is weaker when using the updated profile by \citetalias{Picogna+2019}. 
However, \citetalias{Picogna+2019} argue that at high X-ray luminosities the theory from \citetalias{Owen+2012} breaks down as only the flat region of the $\xi$-$T$ relation (see their section 3.3 for a detailed explanation) is accessible to the X-rays.

\subsection{Model \texttt{Picogna\_10au}}
\label{sec:popsynth_Picogna_10au}

In this model, the planets were inserted at $10\,\au$, which lies outside the location of $7$--$8\,\au$, at which photoevaporation opens a gap. Therefore, planets are now expected to be parked outside of $\Rgap$ again, and indeed a desert of planets in the $\Lx$--$a$-distribution can be observed for this setup. The void is similar in radial extent as the one in the \texttt{Owen\_10au} model, but now encompasses a smaller range of X-ray luminosities ($6\times10^{29}$--$5\times10^{30}\,\ergs$) due to the more vigorous winds resulting from the updated photoevaporation profile. A small pile-up of higher-mass giant planets, comparable to the one observed in model \texttt{Owen\_5au}, can be observed; however, now it includes significantly fewer planets. 
The reason for this is that the insertion location of planets at $10\,\au$ is only marginally larger than the radius at which photoevaporation will open the gap in the \citetalias{Picogna+2019} profile. Therefore, only a small fraction of planets with favourable initial conditions, namely with a high mass and late formation times, can be parked outside of the gap. All remaining planets cross the location of gap opening before photoevaporation becomes dominant.

\subsection{Model \texttt{Picogna\_20au}}
\label{sec:popsynth_Picogna_20au}

Similar to the \texttt{Owen\_20au} model, also here practically no desert of planets caused by photoevaporation is observed anymore. 
Due to the long migration timescales of the planets, most planets are parked outside of $\Rgap$ as photoevaporation can open the gap before the majority of planets can cross this location. 
The number density of planets inside of the void appears to be larger than for the \texttt{Owen\_20au}, possibly due to the higher mass loss rates that disperse the disc more quickly. Therefore, more planets are parked before they reach the inner grid boundary of $0.15\,\au$. Similarly, also for low X-ray luminosities in the $\tclear > 10\,\Myr$ regime more planets are parked across the entire semi-major axis range. Due to the higher mass loss rates, the disc can be depleted more efficiently. Nevertheless, photoevaporation can only weakly impact the final planet parking locations and consequently the planets will still be mostly parked at random locations.

\subsection{Model \texttt{Picogna\_IPMF}}
\label{sec:popsynth_Picogna_random}

The final semi-major axis distribution of planets in the \texttt{Picogna\_IPMF} model is similar to the one in model \texttt{Owen\_IPMF}; however, now the number of planets inside of the observed void is significantly larger. As was seen in the previous models, the photoevaporation profile by \citetalias{Picogna+2019} opens the gap at larger radii, and therefore planets inserted between 10--$20\,\au$ are more likely to cross $\Rgap$ before photoevaporation opens the gap. 

Towards higher $\Lx$, the broad tail of planets appears to have a fuzzier boundary towards lower semi-major axes. As mentioned above, for this photoevaporation profile, weaker winds are expected at higher $\Lx$, meaning that photoevaporation is less efficient in parking the planets in this regime compared to the profile by \citetalias{Owen+2012}. Therefore, the planets' parking location is more strongly dependent on the randomly sampled initial conditions rather than the disc clearing time, and consequently no sharp cutoff of planets can be observed.

\section{Discussion}
\label{sec:discussion}

\subsection{The effect of different photoevaporation profiles}
\label{sec:comp_PE_profiles}

Conceptually, both photoevaporation profiles leave similar imprints in the final $\Lx$--$a$-distribution of giant planets. 
While for $\Lx \lesssim 10^{29}\,\ergs$ giant planet migration is barely affected by disc dispersal via XPE, for $\Lx \gtrsim 5$--$7\times10^{30}\,\ergs$ photoevaporation is the dominant mechanism that stops the inward migration of planets inserted at $5\,\au$ due to the rapid dispersal of the circumstellar material.
For both profiles, an under-density of planets at intermediate values of $\Lx$ can be observed; however its radial extent strongly depends on the insertion location of the planets. 

Only for planet insertion locations of $5\,\au$, major differences between the final $\Lx$--$a$-distribution resulting from the two photoevaporation profiles can be observed. This is because the one by \citetalias{Picogna+2019} opens a gap at larger radii compared to the one by \citetalias{Owen+2012}. Planets inserted at $5\,\au$ are therefore already located inside of $\Rgap$ for the models using the photoevaporation profile by \citetalias{Picogna+2019}.
Nevertheless, besides the pile-up of high-mass planets in the \texttt{Owen\_5au} model, a similarly strong under-density of planets can be observed in both cases, showing that the exact form of the XPE profile does not affect our overall conclusion, namely that for a given range of X-ray luminosities, disc dispersal via XPE will cause a dearth of planets between 1--$10\,\au$.

The profile by \citetalias{Picogna+2019} can be considered as the more realistic one, due to the more accurate treatment of the temperature structure of the disc within the radiation-hydrodynamical calculations. The fact that the observed void of planets for the \texttt{Picogna\_IPMF} model appears to be less strongly confined than for the \texttt{Owen\_IPMF} one suggests, however, that it may be difficult to observe an imprint of XPE within the observed distribution of giant planets. 
In order to investigate this further, we compare the outcomes of our numerical models with actual observational data of exoplanet systems in the following.

\subsection{Comparison with observations}
\label{sec:comp_obs}

\subsubsection{X-ray luminosity sampling}
\label{sec:Lxsampling}

\citet{Owen+2011} argue that the XLF derived from the Taurus cluster is more appropriate as an input to XPE models than the XLF derived for the Orion Nebula Cluster. They argue that due to the removal of flares in the Taurus sample (that is a result of the shorter exposure time of these observations), it can better resemble the quiescent X-ray luminosities of young pre-main-sequence stars. However, stellar flares are a common phenomenon for young stars and need to be accounted for in order to obtain realistic values for stellar X-ray luminosities. The XLF for Orion is based on observations with significantly longer exposure times than the one for Taurus and therefore the stochastic effects of particularly strong X-ray flares are much more washed out by the much longer temporal baseline.

\begin{figure}[h!]
    \centering
    \includegraphics[width=0.9\linewidth]{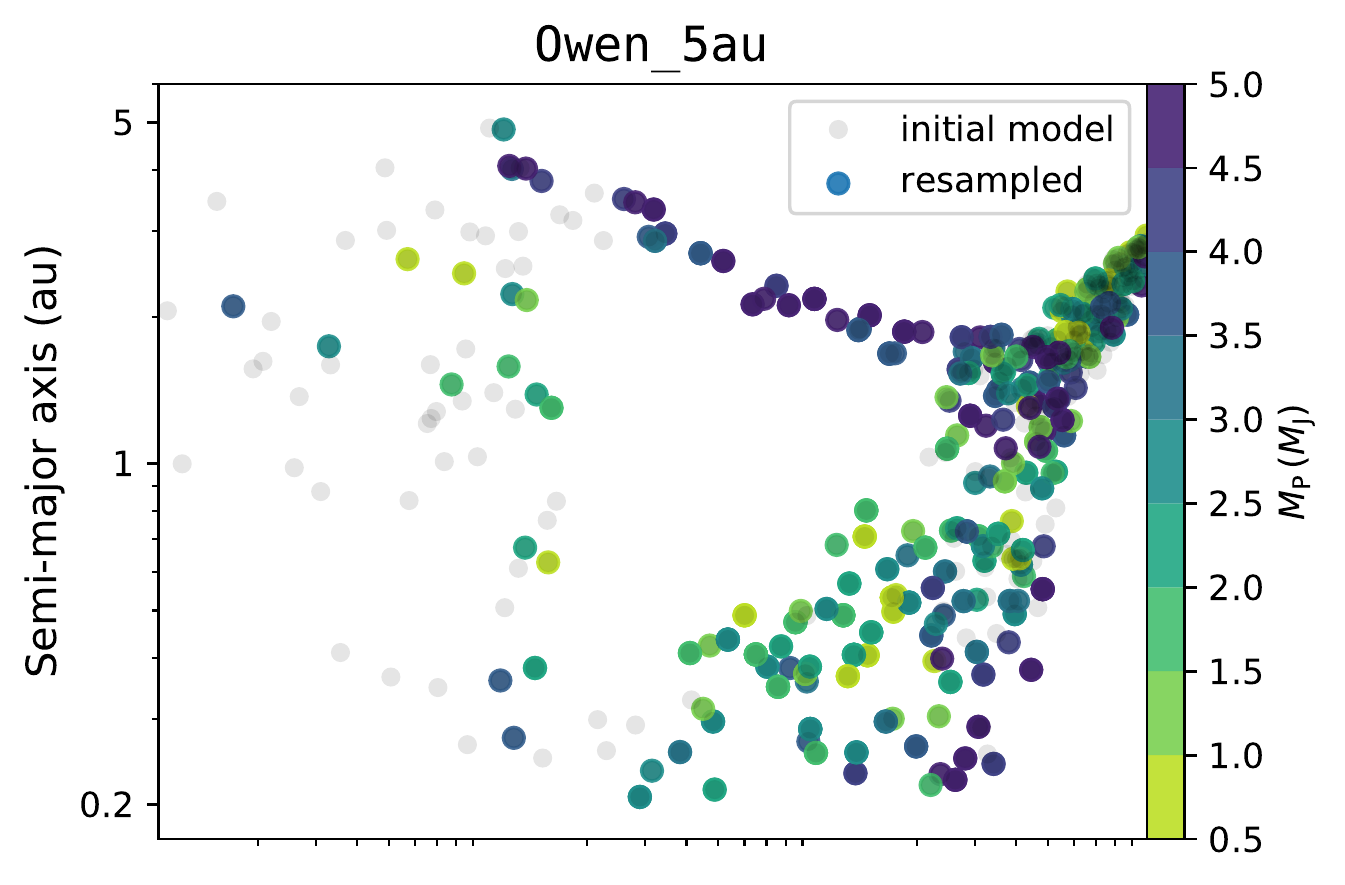}
    \includegraphics[width=0.9\linewidth]{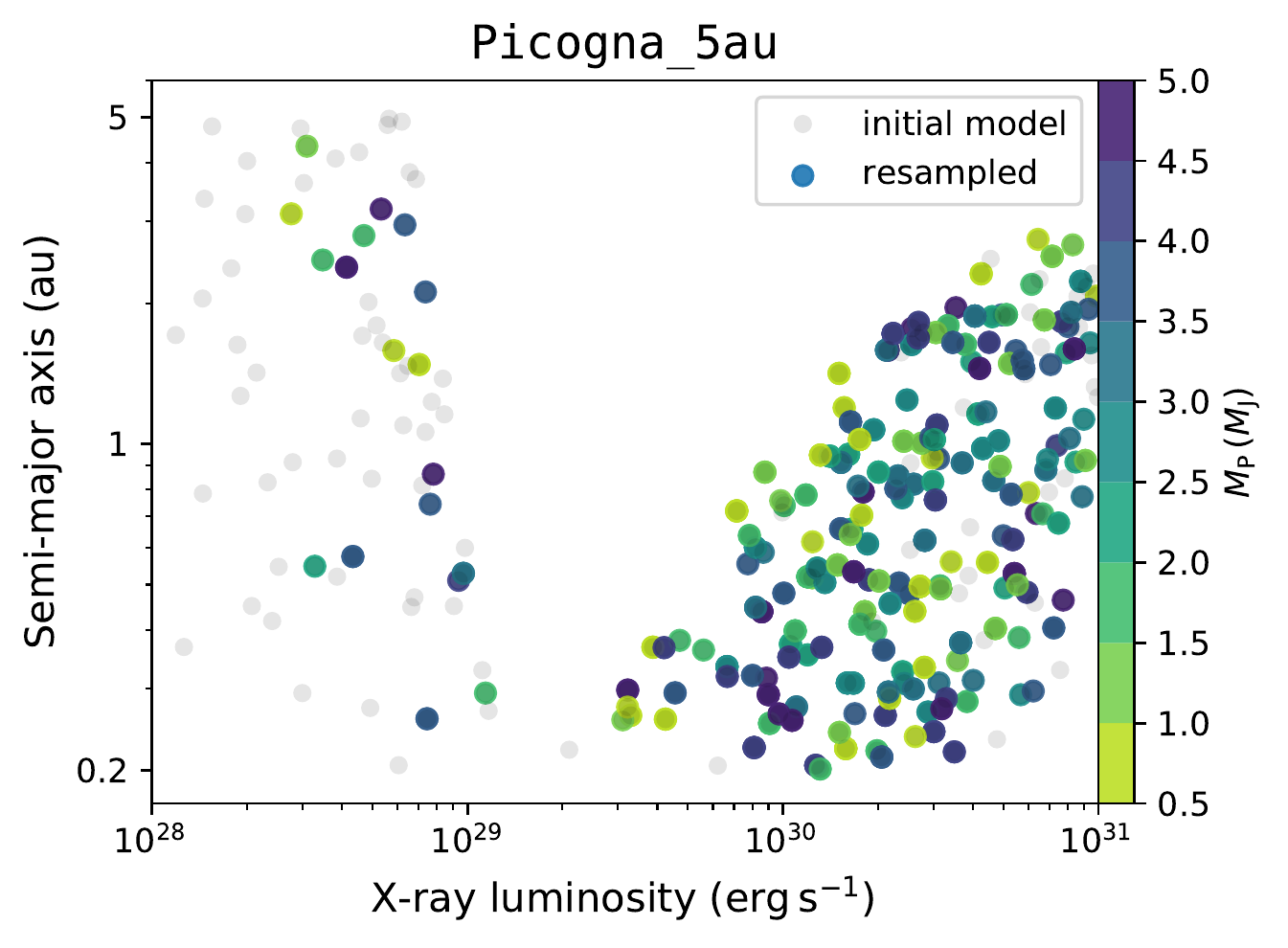}
    \caption{Final $\Lx$--$a$-distribution assuming a realistic sampling of the X-ray luminosity. Grey dots show the initial \texttt{Owen\_5au} and \texttt{Picogna\_5au} models. Based on the XLF of Taurus, each data point was weighted correspondingly and the data were resampled 1500 times, which is shown as the colour-coded sample. Also the pile-up of planets at $0.15\,\au$ were removed from both samples as it corresponds to a numerical artefact rather than a real feature.}
    \label{fig:popsynth_resampled}
\end{figure}

However, as the X-ray luminosity was sampled linearly between $\log(\Lx/\ergs)=27$--$32$ rather than from an observed XLF, the amount of planets residing at the lowest and highest values of $\Lx$ is strongly overestimated in our models.
In order to account for the observational selection function, we resampled the simulated data 1500 times using weights that match the observed XLFs as shown in Fig.~\ref{fig:Lx_func}. The resulting $\Lx$--$a$-distributions of the \texttt{Owen\_5au} and \texttt{Picogna\_5au} models (using the XLF of Taurus) are shown in Fig.~\ref{fig:popsynth_resampled} and are directly compared to the initial models using linear sampling of $\log(\Lx)$. While the number density of points is strongly reduced for $\Lx \lesssim 10^{29}\,\ergs$, the borders of the desert of planets is still well resolved.
Within the limitations of our numerical model, this desert of planets therefore corresponds to a potentially observable feature within the $\Lx$--$a$-distribution of giant planets and their host stars.

\subsubsection{Semi-major axis distribution}
\label{sec:comp_semi}

\begin{figure*}[ht!]
\centering
\includegraphics[width=1.\linewidth]{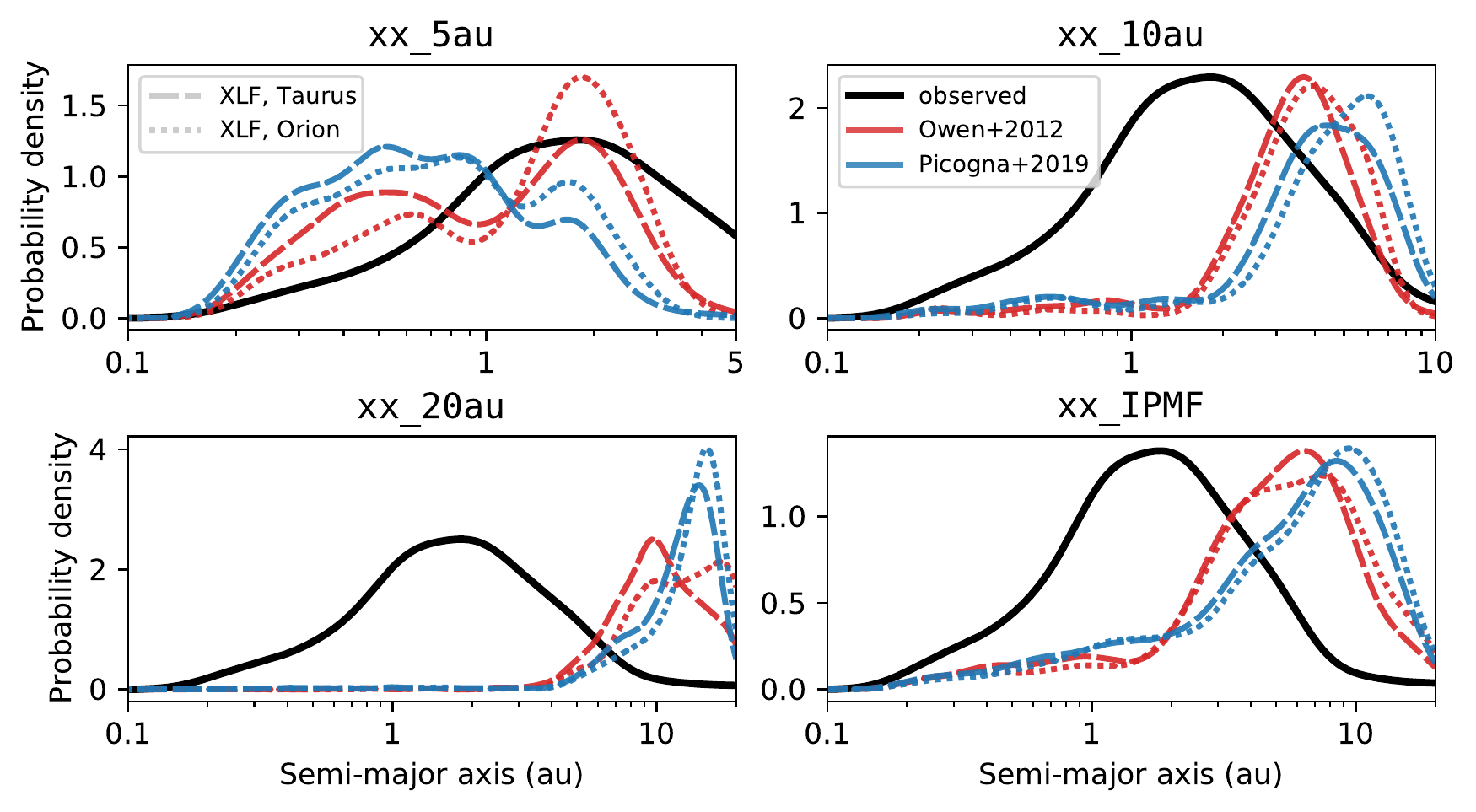}
\caption{Final semi-major axis distributions, calculated using a Gaussian KDE. The blue and red lines correspond to the results from the simulations presented in this study, while the black lines show the observed distribution of giant planets obtained from the NASA Exoplanet Archive. For both datasets, all planets with $a<0.2\,\au$ were removed prior to calculating the KDE. As is described in detail in Sect.~\ref{sec:comp_semi}, the resulting semi-major axis distributions were re-weighted following the XLFs for Taurus and Orion, which are shown in Fig.~\ref{fig:Lx_func}.}
 \label{fig:KDE_comp}
\end{figure*} 

Each panel in Fig.~\ref{fig:KDE_comp} shows the resulting semi-major axis distribution from our population syntheses, obtained using a Gaussian kernel density estimate (KDE) implemented in \texttt{SciPy} \citep{Scipy}, for which we applied Scott's Rule \citep{Scott1992} in order to determine an appropriate bin width for the underlying data. All planets with $a < 0.2\,\au$ were removed from the sample prior to calculating the KDE, in order to remove the numerical artefact at $0.15\,\au$.
To allow an unbiased comparison between the simulations and the observations, we used both XLFs for the following analysis; however, we find that besides in those models, where the planet is inserted at $5\,\au$, no significant difference can be observed in the resulting orbital separations.
This approach therefore ensures that the synthetic semi-major axis distribution can be directly compared to the observed distribution of giant planets\footnote{These data were retrieved from the NASA Exoplanet Archive on 17 December 2020: \url{https://exoplanetarchive.ipac.caltech.edu/}.}, which was scaled to the maximum of the corresponding model using the XPE profile by \citetalias{Owen+2012}, in order to facilitate the readability. To allow a fair comparison between both samples, also for the observed data all planets with $a < 0.2\,\au$ were removed before calculating the KDE.

As previously discussed, the gap opened by photoevaporation is located at different radii in our models, depending on the applied photoevaporation profile. While it lies at $\Rgap = 1$--$2\,\au$ for \citetalias{Owen+2012}, it is located at $\Rgap = 7$--$8\,\au$ for \citetalias{Picogna+2019}. Consequently, one would expect to observe an under-density of planets close to these radii within the orbital distribution of giant planets, and pile-ups of planets closely in- and outside of $\Rgap$. 
Indeed, all of our models show such an under-density of planets in the $\Lx$--$a$-distribution with pile-ups close to the location of gap opening; however their location and extent changes significantly with varying formation location of the planets. This becomes even more apparent in the semi-major axis distribution of the giant planets, which shows that especially the choice of the insertion location of the planets has a dramatic effect on their final parking location.

While the observed sample of giant planets peaks between 1--$3\,\au$, most of the models from our study predict planets to pile up at larger radii, roughly between 3--$10\,\au$.
Only model \texttt{Owen\_5au} successfully reproduces the observed pile-up of giants close to $1\,\au$; however, it over-predicts the amount of planets inside of this location.
While our model is clearly able to produce a pile-up of giant planets, showing its ability in providing a parking mechanism for inward migrating planets, the discrepancy between the observed and the synthetic distribution of planets implies that our numerical model still has some significant caveats. 

In particular the missing self-consistent treatment of planet formation is the biggest caveat of our study as fixed formation locations of the planets leave significantly distinct features in the resulting distribution of gas giants. However, by sampling the planet formation location randomly between 5--$20\,\au$, we are implicitly assuming a constant planet formation efficiency throughout large parts of the planet-forming disc. 
It could be observed in our models that giant planets pile up at larger radii with increasing insertion locations, and that the models in which the planets are inserted at $5\,\au$ (only for \texttt{Owen\_5au}) or $10\,\au$ are more successful in reproducing the observed peak at 1--$2\,\au$, which may hint towards giant planet formation being more likely between 5--$10\,\au$, rather than at radii $>20\,\au$.

This conclusion would be indeed in agreement with observational studies that measured giant planet occurrence rates. For example, by using RV measurements, \citet{Cumming+2008} measured the probability for solar-type stars hosting gas giants (0.3--$10\,\Mjup$) with orbital periods between 2--$2000\,\mathrm{d}$ ($\approx 0.03$--$3\,\au$) to be $10.5\,\%$. However, they proposed a strongly rising giant planet fraction for orbital periods beyond $P \approx 300\,\mathrm{d}$ \citep[$\sim 0.9\,\au$; see also][]{Marcy+2005}, while more recent studies such as the one performed by \citet{Bryan+2016}, which combines RV measurements with direct imaging data, find the giant planet frequency to decline beyond 3--$10\,\au$, therefore suggesting a peak in the giant planet occurrence rate within these radii \citep[see however][who find that the occurrence rate of giant planets plateaus beyond $1\,\au$]{Wittenmyer+2020}. This finding was later confirmed by \citet{Fernandes+2019}, who further combined RV and \kepler\ data to compute unified giant planet occurrence rates for orbital periods of up to $10^4\,\mathrm{d}$. 
However, while our models may suggest a preferential location for giant planet formation, no robust conclusion can be extracted until a self-consistent treatment of planet formation is included into our model.

\subsubsection{$\Lx$--$a$-distribution}
\label{sec:comp_Lxa}

What can be inferred from our models is, however, that photoevaporation is indeed expected to create a desert of planets within the $\Lx$--$a$-distribution of disc-planet systems. Yet, depending on the photoevaporation profile applied, and the assumed insertion location of the planets, its location and size are different, showing the strong dependence of our results on the initial conditions employed in our numerical setup. This strongly limits their predictive power when directly compared to observational data. 

Further, it is important to note the planet distributions presented in Fig.~\ref{fig:popsynth_Owen2012} and Fig.~\ref{fig:popsynth_Picogna2019} are the result of very specific initial conditions in highly idealised models and should be therefore interpreted carefully. Their sole purpose is to investigate if internal XPE can leave a potentially observable imprint in the orbital distribution of giant planets; however, no exact statements about the location or size of such features can be made at this point. 
One reason for this is the previously mentioned missing treatment of planet formation in our model, and therefore the strong dependence on the assumed planet formation locations. 
However, another strong limitation is that the simulations presented in our study were performed at the time of disc dispersal, meaning at ages $< 10\,\Myr$, while extrasolar planets are mainly detected around evolved main-sequence stars with ages of $\sim \mathrm{Gyr}$, with only very few exceptions \citep[e.g. PDS~70,][]{Keppler+2018, Mueller+2018, Haffert+2019}. 
Consequently, any features imprinted in the earliest phases of the disc-planet systems could shift and possibly even be washed out with time due to other processes taking place that are neglected in our approach \citep[e.g. multi-planetary systems and the resulting $N$-body interactions before and after gas-disc dispersal that may possibly even cause outward migration of the planets; see for example][]{Rometsch+2020}. 
Using theoretical arguments, \citet{Monsch+2019} predict that such a void as observed in our population synthesis models, would be expected to shift to lower $\Lx$ with time due to the spin-down of stellar rotation rates with increasing age and the resulting decrease in their magnetic activity, which is tightly linked to the stellar X-ray activity \citep[see e.g.][for reviews]{Guedel2007, BrunBrowning2017}.
Mapping the observed X-ray luminosities of planet-hosting stars to earlier times is non-trivial as the origin of the X-ray emission of late-type stars with spectral types ranging from F to M is still not fully understood. 
While quantitative arguments about the decrease in X-ray luminosity as a function of time can be made \citep[e.g.][and see the discussion in \citet{Monsch+2019}]{GalletBouvier2013, Tu+2015}, it is well beyond the scope of this paper to explicitly calculate the $\Lx$-evolutionary tracks for all stars in our sample. 
At this point, our model is therefore not able to provide the exact location or size of any XPE related features within the $\Lx$--$a$-distribution of giant planets, which would be, however, needed for proving the statistical significance of the void observed by \cite{Monsch+2019}. Certainly, an increase in observational data could help to resolve this issue, and current facilities like eROSITA are expected to soon provide a plethora of X-ray observations of planet-hosting stars. 

Additionally, for a realistic comparison between theoretical models and observational data, it is further necessary to derive realistic occurrence rates from the synthetic planet distributions, which account for selection effects and detection biases introduced by each exoplanet detection technique and/or survey. 
This could be feasible, for example, by using the \texttt{epos}-package developed by \citet{Mulders+2019}, which was successfully tested for the Bern planet population synthesis models. 
Since our model does not treat planet formation, but only investigates how XPE is expected to impact giant planet distributions qualitatively, we refrained from performing any detailed comparisons between simulated and observed giant planet occurrence rates. 
However, in the framework of global population synthesis models, whose goal it is not only to reproduce the exoplanet distributions qualitatively, but also quantitatively, using packages like \texttt{epos} is indispensable. 

\section{Conclusion}
\label{sec:conclusion}

In this paper, we have explored the impact of disc dispersal via XPE onto giant planet migration and focused specifically on how this process can impact the final parking location of giant planets in planetary systems. The main results can be summarised as follows:

\begin{enumerate}
    \item By performing a set of detailed 1D planet population synthesis models with the code \spock, we have found that XPE can indeed create a characteristic void, or under-density, of planets in the semi-major axis versus host star X-ray luminosity plane, as was previously suggested by \citet{Monsch+2019}. 
    By opening an annular gap within the dispersing disc, XPE can provide a parking radius for inward migrating giant planets, so that they pile up both out- and inside of this cavity. 
    \item A comparison between the XPE models by \citet{Owen+2012} with the more recent ones by \citet{Picogna+2019} showed no qualitative difference in the resulting orbital separations of giant planets. However, due to an improved treatment of the underlying kinetic structure of the disc, the latter is more efficient in dispersing the discs, leading to gap opening at larger radii compared to the model of \citet{Owen+2012}, consequently resulting in pile-ups of giants located at larger radii.
    \item The location and especially the size of this desert created by XPE in the $\Lx$--$a$-distribution is strongly dependent on the choice of initial conditions used in our model, specifically the insertion location of the planets. This impedes a direct comparison to the catalogue obtained by \citet{Monsch+2019}, which would need robust measurements of the exact location and size of the void created by XPE in order to confirm that XPE may indeed leave an observational imprint in the observed $\Lx$--$a$-distribution of giant planets.

\end{enumerate}

Our study has shown that XPE could be expected to imprint the final semi-major axis versus host star X-ray luminosity plane, and this may potentially explain the observed pile-up of Jupiter-mass planets close to $\sim 1\,\au$. 
However, with our current models we are unable to make more quantitative statements. Global population synthesis models  including a self-consistent treatment of planet formation and realistic disc dispersal mechanisms are needed in order to get robust results on the $\Lx$--$a$-distribution of giant planet systems that can be directly compared to observations.

\begin{acknowledgements}
    We thank the referee, Richard Alexander, for carefully reading the manuscript as well as his helpful comments that have improved the quality of this paper.
    We acknowledge the support of the DFG priority program SPP 1992 ``Exploring the Diversity of Extrasolar Planets'' (DFG PR 569/13-1, ER 685/7-1) \& the DFG Research Unit ``Transition Disks'' (FOR 2634/1, ER 685/8-1). 
    We further acknowledge the support by the DFG Cluster of Excellence ``Origin and Structure of the Universe''.
    This research has made use of the NASA Exoplanet Archive, which is operated by the California Institute of Technology, under contract with the National Aeronautics and Space Administration under the Exoplanet Exploration Program.
\end{acknowledgements}

\bibliographystyle{aa} 
\bibliography{literature}

\begin{appendix}

\section{Comparison of the different photoevaporation profiles}
\label{sec:AppendixA}

\begin{figure*}[ht!]
    \centering
    \includegraphics[width=\linewidth]{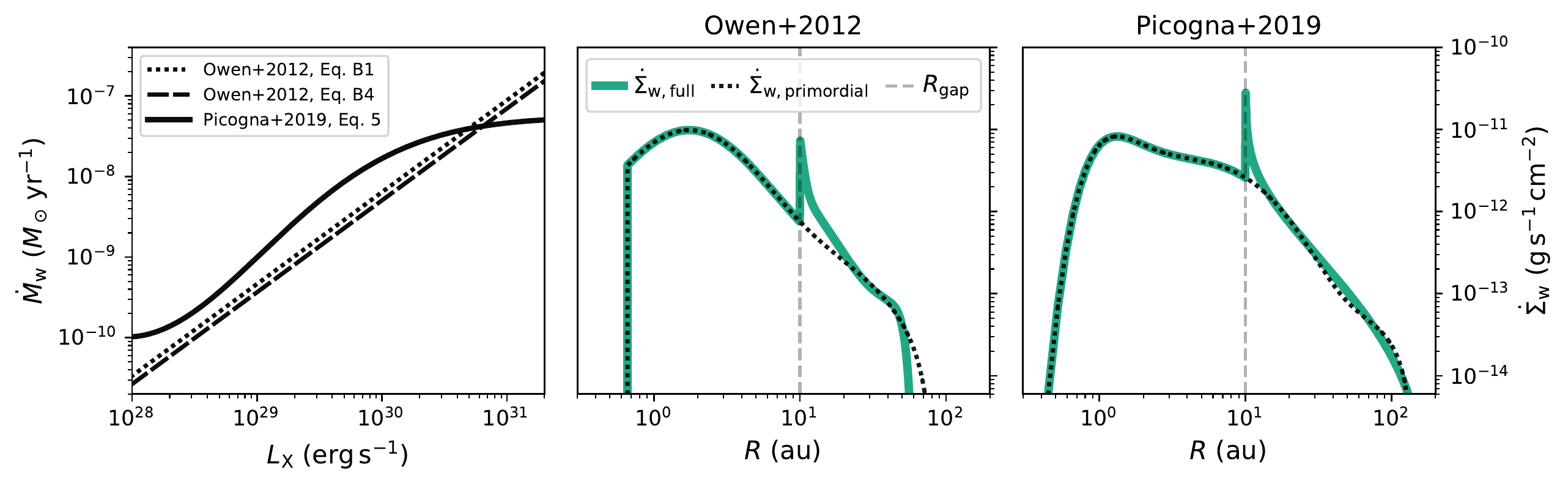}
    \caption{Comparison of the integrated mass loss rates as a function of the X-ray luminosity (left panel) and the surface mass loss profiles as a function of disc radius (centre and right panel) for the photoevaporation profiles by \citet{Owen+2012} and \citet{Picogna+2019}. For the plots showing $\dot{\Sigma}_\mathrm{w, full}$ it is assumed that  photoevaporation already opened a gap in the disc, which has moved up to $10\,\au$, while the inner disc was drained. At this point, the inner edge of the outer disc is located at $10\,\au$, and the column density of the inner disc is less than $2.5\times10^{22}\,\mathrm{cm}^{-2}$. Inside of $10\,\au$ the primordial profile is active (Eq.~\ref{eq:Owen_Sigmadotwind} and Eq.~\ref{eq:Picogna_Sigmadotwind}, respectively), and outside of $10\,\au$ the transition disc profile (Eq.~\ref{eq:Owen_Sigmadotwind_TD} and Eq.~\ref{eq:Picogna_Sigmadotwind_TD}, respectively). The solid lines show the total mass loss profile, while the dotted lines only highlight the primordial profile for each case (i.e. assuming photoevaporation had not opened a gap and the disc is still in its primordial stage).}
    \label{fig:comp_PEprofiles}
\end{figure*}

Fig.~\ref{fig:comp_PEprofiles} shows a comparison of the integrated mass loss rate, $\Mdotwind (\Lx)$, and the surface mass loss profile, $\Sigmadotwind (R)$, from \citet{Owen+2012} and \citet{Picogna+2019}, respectively. 
For the plots showing $\dot{\Sigma}_\mathrm{w, full}$ it is assumed that photoevaporation of a $0.7\,\Msun$ star already opened a gap in the disc, which has moved up to $10\,\au$, while the inner disc had been drained. At this point, the inner edge of the outer disc lies at $10\,\au$, and the column density of the inner disc is less than $2.5\times10^{22}\,\mathrm{cm}^{-2}$, so that the outer disc is directly irradiated by the stellar X-rays. 
Inside of $10\,\au$ the primordial (`diffuse') profile is therefore active (which is valid as long as photoevaporation has not opened a gap yet), and outside of $10\,\au$ the transitional (`direct') profile (which is valid after gap opening). 
The equations plotted in Fig.~\ref{fig:comp_PEprofiles} are explained in the following.

\subsection{Owen et al. (2012)}

In Appendix B1 of \citet{Owen+2012}, the total mass loss rate in primordial discs as a function of X-ray luminosity is described by:

\begin{equation}
\label{eq:Owen_Mdotwind}
    \Mdotwind (\Lx) = 6.25\times10^{-9} \left( \frac{M_\star}{\Msun}\right)^{-0.068} \left( \frac{\Lx}{10^{30}\,\ergs}\right)^{1.14}\,\Msun\,\yr^{-1}.
\end{equation}
The mass loss rate is only weakly dependent on the stellar mass and gives an almost linear scaling with the X-ray luminosity. 
The normalised mass loss profile is given by $\Mdotwind (R)=\int2\pi R\,\Sigmadotwind(R)\,\mathrm{d}R$, where

\begin{eqnarray}
\label{eq:Owen_Sigmadotwind}
    \Sigmadotwind(x>0.7)&\!\!\!\!=\!\!\!\!&10^{(a_1\log(x)^6+b_1\log(x)^5+c_1\log(x)^4)}\nonumber\\&&\times10^{(d_1\log(x)^3+e_1\log(x)^2+f_1\log(x)+g_1)}\nonumber\\&&\times\left(\frac{6a_1\ln(x)^5}{x^2\ln(10)^7} + \frac{5b_1\ln(x)^4}{x^2\ln(10)^6} + \frac{4c_1\ln(x)^3}{x^2\ln(10)^5}\right. \nonumber\\&& + \left.\frac{3d_1\ln(x)^2}{x^2\ln(10)^4}+\frac{2e_1\ln(x)}{x^2\ln(10)^3} + \frac{f_1}{x^2\ln(10)^2}\right)\nonumber\\ &&\times\exp\left[-\left(\frac{x}{100}\right)^{10}\right],
\end{eqnarray}
with $a_1=0.15138$, $b_1=-1.2182$, $c_1=3.4046$, $d_1=-3.5717$, $e_1=-0.32762$, $f_1=3.6064$, $g_1=-2.4918$, and 
    
\begin{equation}
\label{eq:xJ}
    x = 0.85\left(\frac{R}{\au}\right)\,\left(\frac{M_\star}{M_\odot}\right)^{-1}.
\end{equation}
Eq.~\ref{eq:xJ} describes the dimensionless radius (in dependence of the stellar mass) from which on photoevaporation becomes effective, and therefore $\Sigmadotwind(x<0.7)=0$.

In the case of transition discs with inner holes, Eq.~\ref{eq:Owen_Mdotwind} and Eq.~\ref{eq:Owen_Sigmadotwind} change to:

\begin{equation}
\label{eq:Owen_Mdotwind_TD}
    \Mdotwind (\Lx) = 4.8\times10^{-9} \left( \frac{M_\star}{\Msun}\right)^{-0.148} \left( \frac{\Lx}{10^{30}\,\ergs}\right)^{1.14}\,\Msun\,\yr^{-1},
\end{equation}
and

\begin{eqnarray}
\label{eq:Owen_Sigmadotwind_TD}
    \Sigmadotwind(y)&\!\!\!=&\!\!\!\left[\frac{a_2b_2\exp (b_2y)}{R} +\frac{c_2d_2\exp (d_2y)}{R} +\frac{e_2f_2\exp (f_2y)}{R}\right]\nonumber\\
    &&\times\exp\left[-\left(\frac{y}{57}\right)^{10}\right].
\end{eqnarray}
Here, $a_2=-0.438226$, $b_2=-0.10658387$, $c_2=0.5699464$, $d_2=0.010732277$, $e_2=-0.131809597$, $f_2=-1.32285709$, and:

\begin{equation}
    y=0.95\left(R-R_{\rm hole}\right)\left(\frac{M_\star}{M_\odot}\right)^{-1},
\end{equation}
with $\Sigmadotwind(y<0)=0$.

\subsection{Picogna et al. (2019)}

In Sect.~3.1 of \citet{Picogna+2019}, the primordial mass loss rate and surface mass loss profile are defined as:

\begin{equation}
\label{eq:Picogna_Mdotwind}
    \log(\Mdotwind(\Lx)/(M_\odot\, \mathrm{yr}^{-1})) = A_\textrm{L}  \exp{\left[\frac{(\ln{(\log(L_X))}-B_\textrm{L})^2}{C_\textrm{L}}\right]} + D_\textrm{L},
\end{equation}
with $A_\textrm{L} = -2.7326$, $B_\textrm{L} = 3.3307$, $C_\textrm{L} = -2.9868\cdot10^{-3}$, $D_\textrm{L} = -7.2580$, and

\begin{align}
\label{eq:Picogna_Sigmadotwind}
    \Sigmadotwind (R) =& \ln(10)\,\bigg( \frac{6a\ln(R)^5}{R\ln(10)^6} +\frac{5b\ln(R)^4}{R\ln(10)^5} +\frac{4c\ln(R)^3}{R\ln(10)^4} +\frac{3d\ln(R)^2}{R\ln(10)^3} \\ 
    \nonumber &+\frac{2e\ln(R)}{R\ln(10)^2}
    +\frac{f}{R\ln(10)} \bigg)\, \frac{\Mdotwind(R)}{2\pi R}\,M_\odot\,\au^{-2}\,\yr^{-1},
\end{align}
with

\begin{align}
    \Mdotwind(R) =& 10^{(a\log{R}^6 + b\log{R}^5 + c\log{R}^4+ d\log{R}^3)} \\ \nonumber
    &\times 10^{ (e\log{R}^2 + f\log{R} + g)}  \Mdotwind(\Lx),
\end{align}
and $a = -0.5885$, $b = 4.3130$, $c= -12.1214$, $d = 16.3587$, $e = -11.4721$, $f = 5.7248$, and $g = -2.8562$.
For transition discs with inner holes, Eq.~\ref{eq:Picogna_Sigmadotwind} will change to:

\begin{equation}
\label{eq:Picogna_Sigmadotwind_TD}
    \Sigmadotwind(R) = a b^{x} x^{c-1} (x \ln(b)+c)\ \frac{1.12\, \dot{M}(\Lx)}{2\pi R} \ M_\odot\, {\au}^{-2}\, {\yr}^{-1}\,
\end{equation}
where $x=(R-R_\mathrm{gap})$, $a = 0.11843$, $b = 0.99695$ and $c = 0.48835$.

\section{The effect of disc viscosity on the gap location}
\label{sec:viscosity}

\begin{figure}[h!]
    \centering
    \includegraphics[width=\linewidth]{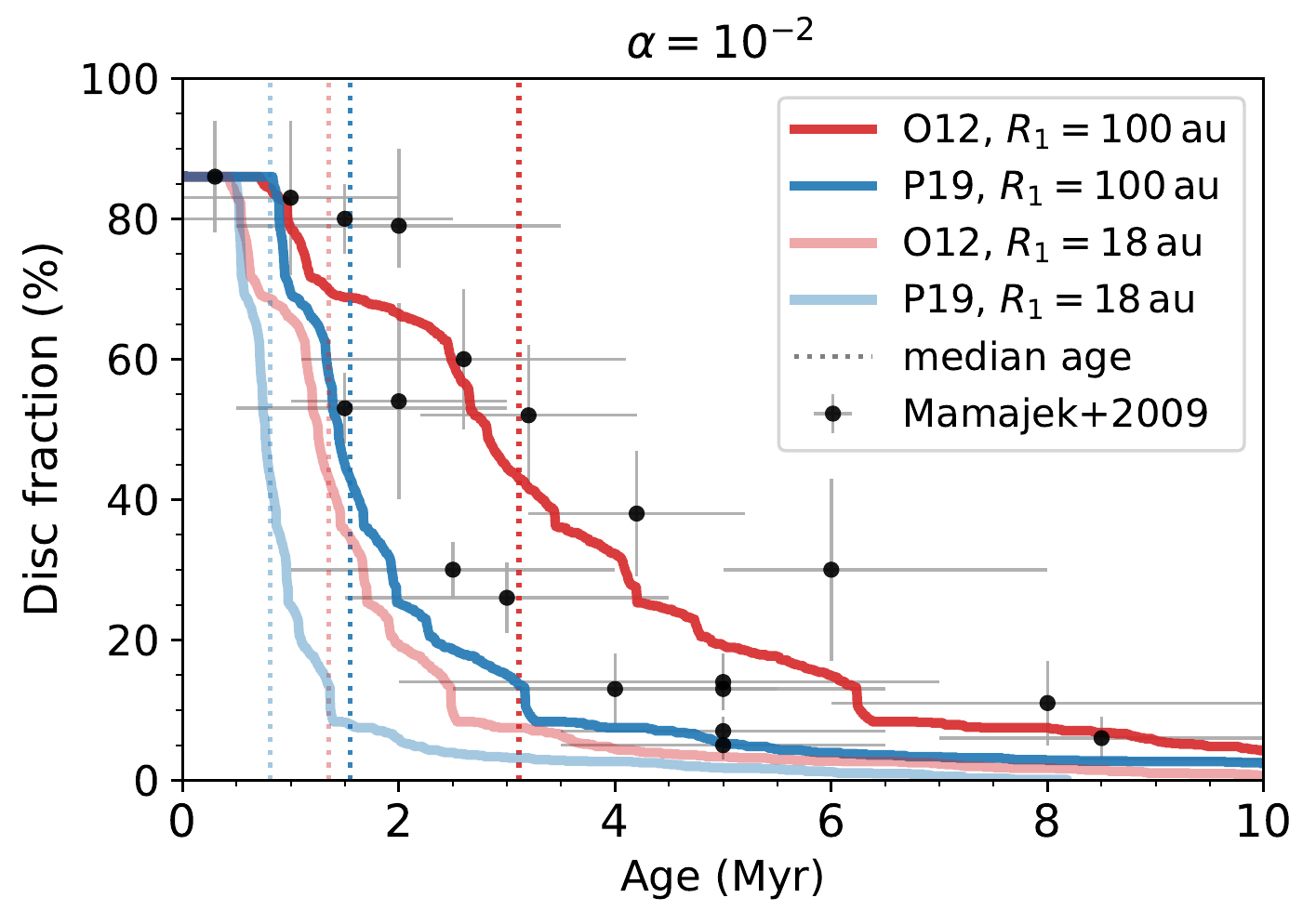}
    \caption{Same as Fig.~\ref{fig:R1_alpha}, but now a fixed viscosity parameter of $\alpha=10^{-2}$ is assumed, while only $R_1$ and the photoevaporation profile are varied within the different runs.}
    \label{fig:R1_alpha1e-2}
\end{figure}

We additionally tested the effect of a larger value for the disc viscosity ($\alpha = 10^{-2}$), which is the same value that was used in previous work by \citet{AP12}. We note, however, that if we kept $R_1 = 18\,\au$ and the remaining initial conditions of the discs in our model the same, unreasonably short total disc lifetimes would be obtained. 
Therefore we performed the same test as previously described in Sect.~\ref{sec:1dpopsynth} in order to obtain an appropriate combination of $R_1$ and $\alpha=10^{-2}$ that matches observed disc fractions as a function of cluster age. The results are shown in Fig.~\ref{fig:R1_alpha1e-2}. 
While \citet{AP12} considered EUV-dominated winds in their disc models, XPE yields more than two orders of magnitude higher wind mass loss rates. Combined with the stronger accretion onto the host star due to the higher viscosity, this would result in too short median disc lifetimes of $\lesssim 1\,\Myr$ if $R_1=18\,\au$ is used in our models. 
In combination with higher disc viscosities of $\alpha=10^{-2}$, we found $R_1 = 100\,\au$ to yield more realistic median disc lifetimes of 1--$3\,\Myr$.

\begin{figure}[h!]
    \centering
    \includegraphics[width=0.9\linewidth]{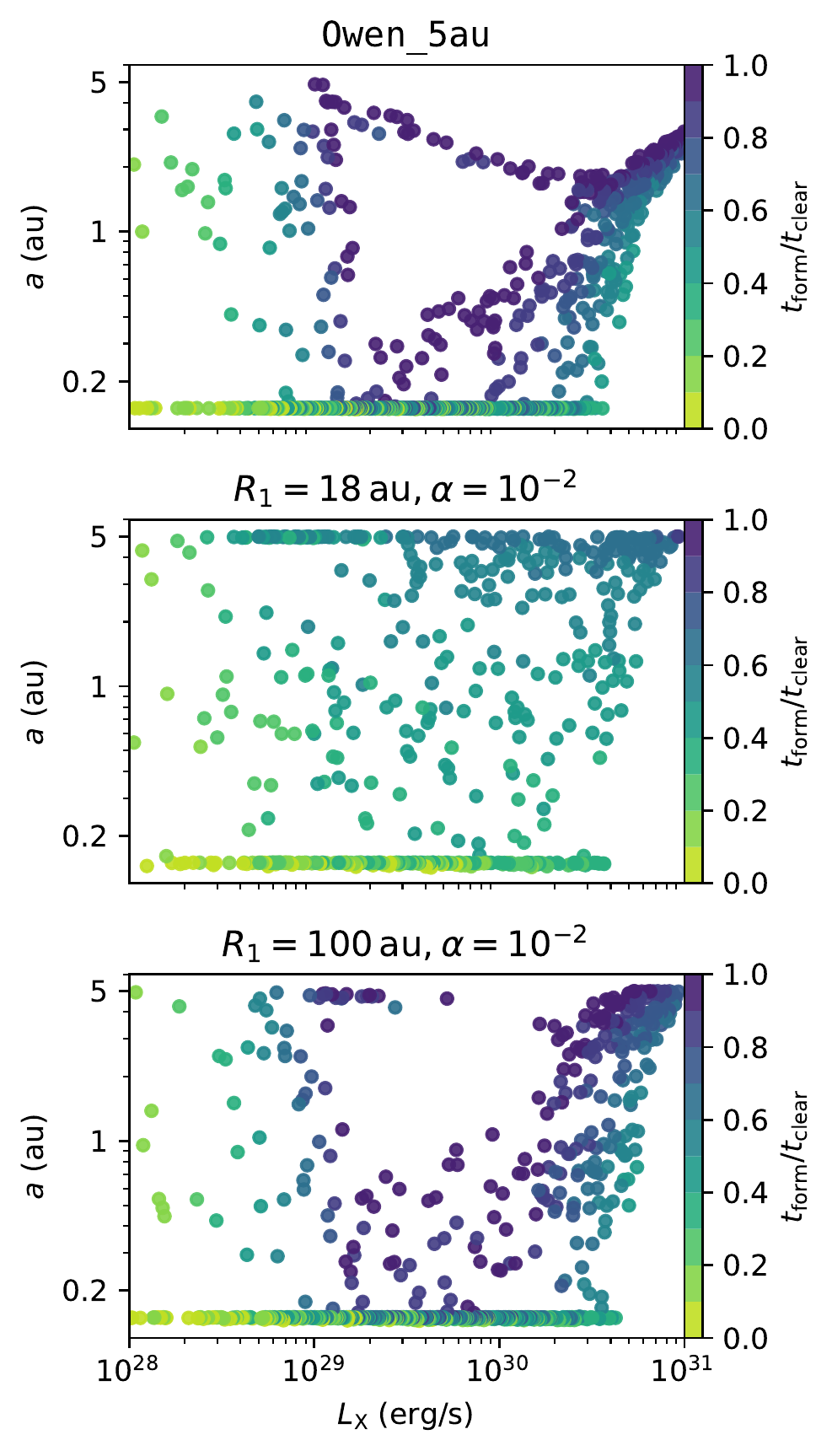}
    \caption{Comparison of the resulting $\Lx$--$a$-distributions using different values for $\alpha$ and $R_1$. The top panel shows the \texttt{Owen\_5au} using the default values of $\alpha=6.9\times10^{-4}$ and $R_1=18\,\au$, while in the middle and lower panels $R_1=18\,\au, \alpha=10^{-2}$ and $R_1=100\,\au, \alpha=10^{-2}$ were used.}
    \label{fig:comp_alpha}
\end{figure}

Using these two sets of $R_1$ and $\alpha$, we reran the \texttt{Owen\_5au} models in order to test the effect of the viscosity parameter on the desert of planets that is observed in the $\Lx$--$a$-distribution of giant planet systems. 
The results are shown in Fig.~\ref{fig:comp_alpha}. 
For the case of $R_1=18\,\au$ and $\alpha=10^{-2}$, the desert of planets that could be observed in the $\Lx$--$a$-distribution of the \texttt{Owen\_5au} model, has now mostly disappeared. 
Only towards $\Lx \lesssim 10^{29}\,\ergs$ can a small desert of planets close to the insertion location of $5\,\au$ can be seen. 
This is, however, to be expected as with $R_1=18\,\au$ most of the disc mass is concentrated relatively close to the host star. Combined with a large value of $\alpha=10^{-2}$ for the disc viscosity, this would yield a viscous timescale of $4.8\times 10^{4}\,\yr$ at $R_1$, leading to the significantly faster accretion of the disc material onto the host star than compared to the \texttt{Owen\_5au} model, for which $\tnu=7\times 10^{5}\,\yr$. 
In this model, the discs would be accreted so quickly that already smaller X-ray luminosities of the star would suffice in order to disperse the discs via XPE, explaining why the void of planets has shifted towards lower values of $\Lx$. 

In contrast, for $R_1=100\,\au$ and $\alpha=10^{-2}$ a clearly confined desert of planets can be observed again. While it is located at approximately the same range of X-ray luminosities as in the \texttt{Owen\_5au} model, its radial size is slightly decreased. This can be related to the different properties of the disc (e.g. different mass at a given radius and different viscosity) and the resulting different migration speed of the individual planets.
Due to the higher viscosity, more planets will end up at the inner grid and possibly end up as hot Jupiters. Nevertheless, the effect of higher-mass planets being parked outside of the desert can also be observed for higher $\alpha$ in this case, even though with significantly reduced number density. 

In conclusion, the evidence for an XPE-related desert of planets within the $\Lx$--$a$-distribution of giant planets is not sensitive to the exact choice of the disc viscosity as long as the models reproduce the observed disc lifetimes in combination with viscous accretion and XPE.

\section{The effect of planet accretion on the gap location}
\label{sec:planet_accretion}

In our 1D model, the mass-flow of the gap-crossing material is modelled following the prescription derived by \citet{VerasArmitage2004}, which is based on 2D hydrodynamical simulations performed by \citet{Lubow+1999} and \citet{D'Angelo+2002}:

\begin{equation} 
 \frac{\epsilon}{\epsilon_{\rm max}}  \simeq 
 1.668 \left(\frac{\Mpl}{\Mjup}\right)^{1/3} 
 \exp \left(-\frac{\Mpl}{{1.5\Mjup}} \right) + 0.04.
 \label{eq:leakage}
\end{equation} 
Here, $\epsilon=\dot{M}_\mathrm{p}/\dot{M}_\mathrm{disc}$ describes the efficiency of mass accretion across the planetary gap that is, the ratio of the planetary accretion rate to the viscous disc accretion rate at large radii \citep{VerasArmitage2004}. The parameter $\epsilon_\mathrm{max}$ is an adjustable parameter that can be used to test the results' dependence on the efficiency of planetary accretion. In this work it was set to $\epsilon_\mathrm{max}=0.5$ to enable a direct comparison to previous work \citep{AP12, ER15, Jennings+2018}.

\begin{figure}[h!]
    \centering
    \includegraphics[width=0.87\linewidth]{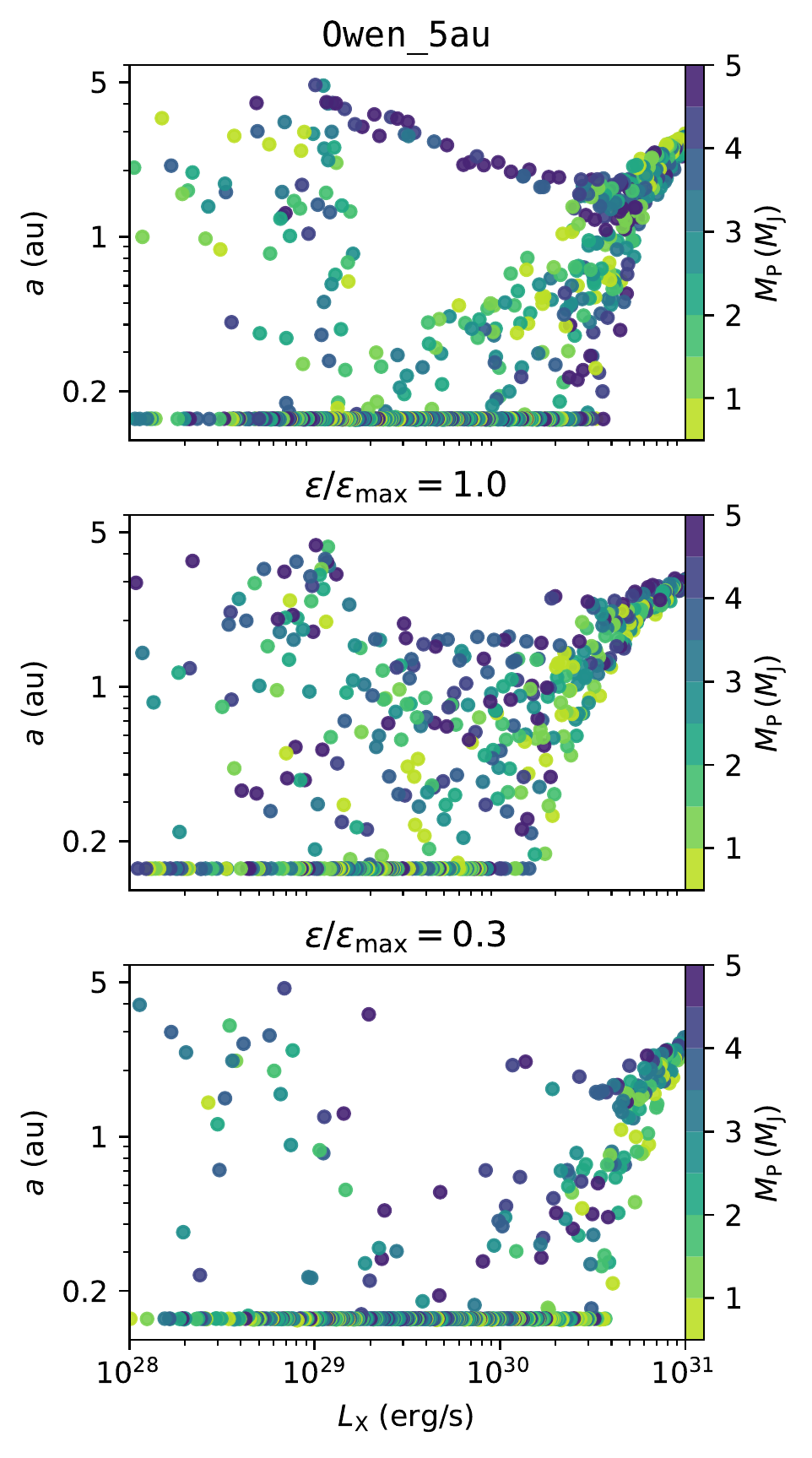}
    \caption{Comparison of the resulting $\Lx$--$a$-distributions using different values for the planetary accretion efficiency. The top panel shows the default model using Eq.~\ref{eq:leakage}, while in the other two models, it was set to a constant value of $\epsilon/\epsilon_\mathrm{max}=1.0$ (centre) and $\epsilon/\epsilon_\mathrm{max}=0.3$ (bottom). The colour-coding represents the planet mass. }
    \label{fig:comp_accretion}
\end{figure}

\citet{AP12} argue, however, that the accretion efficiency has the biggest impact on setting the final parking location of the planets in their model. 
We therefore ran different models based on the \texttt{Owen\_5au} one, in which the accretion efficiency was set to a constant value of $\epsilon / \epsilon_{\rm max}=1$ and $\epsilon / \epsilon_{\rm max}=0.3$ as was tested by \citet{AP12}. The resulting $\Lx$--$a$-distributions are shown in Fig.~\ref{fig:comp_accretion}. 
For $\epsilon / \epsilon_{\rm max}=1$ (i.e. $\Mpl \approx 0.5\,\Mjup$), the void created by XPE shifts towards larger radii and has mostly disappeared due to the insertion of planets at $5\,\au$. In contrast to the \texttt{Owen\_5au} model, there is no significant difference in the radial distribution for planets of different masses, and the strong bifurcation between lower- and higher-mass giants that was observed previously has entirely disappeared. 
For $\epsilon / \epsilon_{\rm max}=0.3$ (i.e. $\Mpl \approx 2.5\,\Mjup$), most planets end up at the inner boundary. However, the desert of planets observed in the \texttt{Owen\_5au} model can still be weakly surmised. 
This confirms that also our model is strongly dependent on the underlying planetary accretion prescription. Consequently, detailed hydrodynamical calculations of the accretion process of planets embedded in photoevaporating discs are required to provide more realistic prescriptions that can be implemented into 1D planet population synthesis approaches.

In contrast to the strong dependence on the planetary accretion prescription, \citet{AP12} only find a weak dependence on the planet mass and especially on their insertion location (i.e. the underlying assumption on where and when planets form).
The latter conclusion is in contradiction with our results, but the discrepancy can be readily understood as their model only treats EUV-driven photoevaporation, for which the mass loss is mostly concentrated around the gravitational radius (i.e. $\sim 1\,\au$). Consequently the surface density at larger radii will be less strongly depleted compared to a disc irradiated by X-rays, for which the mass loss extends to much larger radii.
In a purely EUV-irradiated disc, the final parking location of a planet inserted at $5\,\au$ or $10\,\au$ is therefore solely set by its unperturbed type~II migration rates, which mostly depend on the local disc surface density and viscosity. Only once the planets reach the innermost parts of the disc, their migration tracks will be directly affected by photoevaporation.
In contrast, planets embedded in a disc irradiated by X-ray dominated winds will be subject to weaker gas torques already at the time of their formation, and therefore be parked at larger radii compared to a model assuming EUV-dominated winds.

\end{appendix}
\end{document}